\documentclass[12pt]{article}

\usepackage{amssymb}
\usepackage{amsmath}
\usepackage[pdftex]{graphicx}
\usepackage{float}
\usepackage{enumitem}
\usepackage{multicol}
\usepackage{sidecap}
\usepackage{verbatim}
\usepackage{vmargin}
\everymath{\displaystyle}
\usepackage{color}
\usepackage{subfigure}

\usepackage{array}
\newcolumntype{C}[1]{>{\centering\let\newline\\\arraybackslash\hspace{0pt}}m{#1}}

\setpapersize{USletter}
\setmarginsrb{1in}{1in}{1in}{1in}{0pt}{0mm}{0pt}{15mm}

\begin{document}

\begin{titlepage}
\bigskip \begin{flushright}
\end{flushright}
\vspace{1cm}
\begin{center}
{\Large \bf {Five-Dimensional Eguchi-Hanson Solitons in Einstein-Gauss-Bonnet Gravity}}\\
\end{center}
\begin{center}
{A. W.C. Wong\footnote{E-mail:  a57wong@uwaterloo.ca} and R. B. Mann\footnote{E-mail: rbmann@perimeterinstitute.ca}}\\
{Department of Physics \& Astronomy\\
 University of Waterloo, Waterloo, ON N2L 3G1, Canada\\}
 \vspace{1cm}
\end{center}

\begin{abstract}
\noindent Eguchi-Hanson solitons are odd-dimensional generalizations of the four-dimensional Eguchi-Hanson metric and are asymptotic to AdS$_5$/$\mathbb{Z}_p$ when the cosmological constant is either positive or negative.  We find soliton
solutions to Lovelock gravity in 5 dimensions that are generalizations of these objects. 
\end{abstract}
\end{titlepage}

\section{Introduction}

The AdS/CFT correspondence has played a pivotal role in
 modern theoretical physics for more than a decade because of its conceptual importance and its broad applications \cite{Maldacena1999-AdSCFT}. 
It conjectures that for some gravitational theory on a $D$-dimensional spacetime of the form $\text{AdS}_d \times \text{M}_{D-d}$ where M is a compact space of $D-d$ dimensions ($D=10$ for string-theory and $D=11$ for M-theory)
then there also exists  a mathematically equivalent $(d-1)$-dimensional Conformal Field Theory (CFT).  
As a consequence, solutions to the vacuum Einstein equations in $(d-1)$-dimensions have particular importance since they furnish both ground states for string theory and background manifolds for the dual CFT.

An interesting case in point are generalizations of the Eguchi-Hanson (EH) metric.  EH metrics are solutions of
the four-dimensional  vacuum Euclidean Einstein equations, and can be regarded as a special case of the Atiyah-Hitchen metric.  This latter metric can be embedded in M-theory to furnish new M2- and M5-brane solutions \cite{Ghezelbash2004-AtiyahHitchin}.   Odd-dimensional generalizations of the formerly 4-dimensional EH metric were  discovered a few years ago in Einstein gravity and are referred to as Eguchi-Hanson solitons. Originally obtained in 5 dimensions upon
taking a particular limit of Taub-NUT space \cite{Clarkson2006-5DSoliton}, they can also be obtained from a consideration
of static regular bubbles with AdS asymptotic behaviour \cite{Copsey2007-Bubbles2}  and 
generalize to any odd-dimensionality  \cite{Clarkson2006-OddSoliton}.  These latter cases can be derived from a set of inhomogeneous Einstein metrics on sphere bundles fibred over Einstein-Kahler spaces \cite{Page1987-Inhomogeneous,Lu2004-Inhomogeneous}.

EH solitons are asymptotic to AdS$_d$/$\mathbb{Z}_p$ ($p \ge 3$) and have Lorentzian signature. As such they furnish 
interesting non-simply connected background manifolds for the CFT boundary theory. Perturbatively
the EH soliton has the lowest energy in its topological class, suggesting that it is either a new ground state in string theory or that some as yet unknown ground state exists.

Quantum gravitational corrections generally induce higher-order curvature terms in the low-energy effective action \cite{Ogawa2011-EGBAdS}.  
When Einsteinian gravity is considered in higher dimensions $n \ge 4$, the applicability of the theory still stands but it is not sufficient in realizing the free-field dynamics involved \cite{Dadhich2005-GB}. Consequently much attention to this end has concentrated on Lovelock gravity \cite{Lovelock1971-LG1,Lovelock1970-LG2,Lovelock1975-LG3}, since its field equations are always of 2nd order, even though the associated action is non-linear in the curvature tensor.    The most commonly studied special case of Lovelock gravity is Einstein-Gauss-Bonnet (EGB) gravity, which only contains   curvature-squared terms, and implications for AdS/CFT of incorporating this term have recently been carried out \cite{Ogawa2011-EGBAdS}.   

In this paper we   investigate the existence of   EH soliton solutions in EGB gravity in five dimensions.  We find that
generalizations of the 5-dimensional EH soliton  \cite{Clarkson2006-5DSoliton} do indeed exist.  We construct such solutions via   semi-analytic  and numerical methods.  These solutions smoothly approach the Einsteinian EH soliton solutions in the limit that
the Gauss-Bonnet parameter $\alpha\to 0$.
 
The outline of our paper is as follows.
In Section \ref{sec:EH_soliton}, we will introduce the reader of the EH soliton.
In Section \ref{sec:EGB_gravity}, the action formalism of EGB gravity will be introduced and EGB field equations will be shown in Einstein-notation.
Using a similar notation to \cite{Clarkson2006-5DSoliton,Clarkson2006-OddSoliton} for the EH soliton, we transform the metric
to  a more convenient form in Section \ref{sec:EGB_gravity}, and also present
the explicit EGB field equations scaled in dimensionless units. We then compute large$-r$ and small-$r$ 
 power-series solutions to the equations in Section \ref{sec:semianalytical_results}, and discuss  the required metric regularity conditions for a soliton solution. The full numerical results of solution are given in Section \ref{sec:numerical_results}.
Finally, Section \ref{sec:conclusions_and_discussions} contains a discussion of our results.

Throughout this paper, we will use both greek and latin characters for upper and lower indices of the tensors but reserve latin characters for 'dummy' variables.
These indices will run through $n$ variables where $n=5$ is the dimensionality of the spacetime.

\section{The Eguchi-Hanson Soliton}
\label{sec:EH_soliton}

In this section we briefly review the structure and properties of the EH soliton. Its metric
has the form
 \cite{Clarkson2006-5DSoliton}
\begin{equation}
ds^2 = 
-g(r) dt^2 + \frac{r^2f(r)}{4} [ d\psi + \cos \theta d\phi ]^2 + \frac{dr^2}{f(r)g(r)} + \frac{r^2}{4} d\Omega_2^2
\label{eqn:EH5D_metric}
\end{equation}
where $d\Omega_2^2 = [d\theta^2 + \sin^2\theta d\phi^2]$ is the unit 2-sphere and
\begin{equation}
\begin{array}{ccc}
 g(r) = 1+\frac{r^2}{\ell^2}, & f(r) = 1-\frac{a^4}{r^4}, & \Lambda = -\frac{6}{\ell^2}
\end{array}
\label{eqn:EH5D_metric_fg}
\end{equation}
The EH soliton is an exact solution to the 5-dimensional Einstein equations. 
In the $\ell \rightarrow \infty$ limit, the 5D EH soliton \eqref{eqn:EH5D_metric} recovers the EH metric
\begin{equation}
ds^2 = 
\frac{r^2f(r)}{4} [ d\psi + \cos \theta d\phi ]^2 + \frac{dr^2}{f(r)} + \frac{r^2}{4} d\Omega_2^2
\label{eqn:EH4D_metric}
\end{equation}
on a constant  $t$ hypersurface. 

In general the metric \eqref{eqn:EH5D_metric} will not be regular unless certain conditions are imposed on
its parameters.  As $r\rightarrow a $, regularity in the $\left( r,\psi \right) $ section implies that $\psi $ has
period $2\pi/\sqrt{g(a)}$. Removing  string singularities at $\left( \theta =0,\pi \right) $ implies that an
integer multiple of this quantity must equal $4\pi $. Consequently 
\begin{equation}
a^{2}=\ell ^{2}\left( \frac{p^{2}}{4}-1\right)   \label{EHmatch}
\end{equation}%
where  $p\geq 3$ is an integer, implying that $a>\ell $. The
metric \eqref{eqn:EH5D_metric} can be rewritten as
\begin{equation}
ds^{2}=-\left( 1+\frac{r^{2}}{\ell ^{2}}\right) dt^{2}+\frac{r^{2}}{p^{2}}%
f(r)\left[ d\psi +\frac{p}{2}\cos (\theta )d\phi \right] ^{2}+\frac{dr^{2}}{%
\left( 1+\frac{r^{2}}{\ell ^{2}}\right) f(r)}+\frac{r^{2}}{4}d\Omega
_{2}^{2}~~~ \label{mtrcreg}
\end{equation}%
where now
\begin{equation*}
f(r)=1-\frac{\ell ^{4}}{r^{4}}\left( \frac{p^{2}}{4}-1\right) ^{2}
\end{equation*}
and we have rescaled $\psi$ so as to have period $2\pi$.

The regularity condition (\ref{EHmatch}) implies that the metric   \eqref{mtrcreg}
is asymptotic to AdS$_5$/$\mathbb{Z}_p$ where $p\geq 3$. 
The AdS/CFT correspondence conjecture 
states that string theory on spacetimes that asymptotically
approach AdS$_{5}$\ $\times $ $S^{5}$ is equivalent to a conformal field theory
(CFT) ($N=4$\ super Yang-Mills $U(N)$ gauge theory) on its boundary $\left(
S^{3}\times \mathbb{R}\right) \times S^{5}$.   Here the boundary of the EH soliton
is a quotient of AdS,  implying the existence of extra light states in the gauge theory. 
The density of low energy states is not
affected even though the volume of $S^{3}$ has been reduced to $S^{3}/\mathbb{Z}_p$.

 As noted in the introduction, the EH soliton perturbatively has the lowest energy in its topological class.
In the context of string theory, the implications are either \cite{Clarkson2006-OddSoliton}:

\begin{enumerate}
\item If the soliton is non-perturbatively stable, then it is a new ground state to string theory
\item Otherwise, if the soliton is non-perturbatively unstable, then there exists another unknown ground state to string theory
\end{enumerate}
Moreover, the perturbed solutions from the EH soliton that are still asymptotic to AdS$_5$/$\mathbb{Z}_p$ ($p \ge 3$) have higher energies than the EH soliton which suggests its position in a energetically local minimum.

\section{Einstein-Gauss-Bonnet Gravity}
\label{sec:EGB_gravity}

The unique combination of curvature-squared terms that is contributed by Gauss-Bonnet gravity towards the Einstein-Hilbert action is 
\begin{equation}
\mathcal{L}_{\text{GB}} = R_{abcd}R^{abcd} - 4R_{ab}R^{ab} + R^2
\label{eqn:GB_action_terms}
\end{equation}
So that the total action $\mathcal{I}_{\text{EGB}}$ becomes
\begin{equation}
\mathcal{I}_{\text{EGB}} = \frac{1}{16 \pi \text{G}} \int d^n x \sqrt{-g} \left[ -2 \Lambda + R + \alpha \mathcal{L}_{\text{GB}} \right]
\label{eqn:EGB_action}
\end{equation}
where $n$ is the dimensions of the metric and $\alpha$ is the Gauss-Bonnet coefficient for controlling the relative magnitude of the contributing curvature-squared terms.
If the action from \eqref{eqn:EGB_action} is varied with respect to the metric (and neglecting boundary terms) we obtain the EGB field equations
 \begin{equation}
\begin{array}{lcll}
\Bigl[ R_{\mu \nu} - \frac{1}{2}Rg_{\mu \nu} + \Lambda g_{\mu \nu} \Bigr] & - &
\alpha \Bigl[ \ \frac{1}{2}g_{\mu \nu}(R_{abcd}R^{abcd} - 4R_{ab}R^{ab} + R^2) - 2RR_{\mu \nu} & \\ & & \ \ \ \ \ + 4R_{\mu a}R^{a}_{\nu} + 4R^{a b}R_{\mu a \nu b} - 2R^{abc}_{\mu}R_{\nu a b c} \ \ \ \ \ \ \ \ \ \Bigr] = 0 &
\end{array}
\label{eqn:EGB_fieldequations}
\end{equation}
Eqn \eqref{eqn:EGB_fieldequations} shows explicitly the contribution of the Gauss-Bonnet terms towards the pure Einstenian field equations.

\noindent In order to search for EH-type soliton solutions to the equations  \eqref{eqn:EGB_fieldequations}, we employ the
ansatz
\begin{equation}
ds^2 = 
- \frac{r^2}{\ell^2} g(r) dt^2 + \frac{r^2f(r)}{4} [ d\psi + \cos \theta d\phi ]^2 + \frac{\ell^2}{r^2} \frac{dr^2}{f(r)h(r)} + \frac{r^2}{4} d\Omega_2^2
\label{eqn:EH5D_our_metric}
\end{equation}
and require that   $\{ f(r),g(r),h(r) \}$ each have an asymptotic limit of $1$.  In the limit $\alpha \to 0$,
\begin{equation}
\begin{array}{cccc}
g(r) = 1+\frac{\ell^2}{r^2}, & h(r) = 1+\frac{\ell^2}{r^2}, & f(r) = 1-\frac{a^4}{r^4}, & \Lambda = -\frac{6}{\ell^2}
\end{array}
\label{eqn:EH5D_our_metric_fgh}
\end{equation}
recovering the metric \eqref{eqn:EH5D_metric}.

The EGB field equations \eqref{eqn:EGB_fieldequations} can be written explicitly into a set of ordinary differential equations using \eqref{eqn:EH5D_our_metric}.  We obtain 7 non-trivial field equations $\{ \mathcal{E}_{tt},\mathcal{E}_{rr},\mathcal{E}_{\psi\psi},\mathcal{E}_{\theta\theta},\mathcal{E}_{\phi\phi},\mathcal{E}_{\psi\phi},\mathcal{E}_{\phi\psi} \}$. 
that in turn can be reduced to 3 independent  field equations $\{ \mathcal{E}_{rr},\mathcal{E}_{\psi\psi},\mathcal{E}_{\theta\theta} \}$. Rescaling the parameters and variables in the field equations via 
\begin{equation}
\begin{array}{ c c c c c }
x = \frac{r}{\ell}, & f_{xx}(x) = \ell^2f_{rr}(r), & f_x(x) = \ell f_r(r), & \Lambda_{*} = \ell^2 \Lambda, & \alpha_{*} = \frac{\alpha}{\ell^2}
\end{array}
\label{eqn:dimensionless_parameters}
\end{equation}
where the subscript denotes the derivative with respect to the relevant variable, 
we find
\begin{equation}
\begin{array}{l}
\Big[
\ 12f^2h + 12x^4f_xh_xfh - 64fh - 32h_xf + 8xhf^2 - 128f_xh + 24x^2f_x^2h -16x^2f_xh_x \\
\ \ - 32x^2f_{xx}h + 8x^4f_x^2h^2 + 48x^2f^2h^2 + 24x^2f_{xx}fh + 12x^2f_xh_xf +80xf_xfh \\ 
\ \ + 8x^4f_{xx}fh^2 + 64x^3f_xfh^2 + 24x^3h_xf^2h 
\Big] \alpha_* 
\\ \ \ \ \ \ \ \ \ \ \ \ \ \ \ \ \ \ \ \ \ \ \ \ \ \ \ \ \ \ \ \ + \\
\Big[
\ 16 - 24x^2fh - 16x^3f_xh - x^4f_xh_x - 6x^3h_xf - 2x^4f_{xx}h - 4f - 4\Lambda_*x^2
\Big]
= 0
\end{array}
\label{eqn:EGB_fieldequation_tt}
\end{equation}
\begin{equation}
\begin{array}{l}
\Big[
\ - 48f^2g^2h + 32xfh_xg^2 - 24xf^2h_xg^2 - 16x^2fg_x^2h + 32xf_xg^2h - 48x^2f^2g^2h^2 \\
\ \ + 64fg^2h + 4x^4f^2g_x^2h^2 - 24xff_xg^2h - 24x^3ff_xg^2h^2 + 12x^2f^2g_x^2h + 32x^2fg_{xx}gh \\
\ \ - 72xf^2g_xgh + 96xfg_xgh - 8x^4f^2g_{xx}gh^2 + 16x^2f_xg_xgh - 40x^3f^2g_xgh^2 \\
\ \ - 24x^3f^2h_xg^2h - 12x^2f^2g_xh_xg + 16x^2fg_xh_xg - 12x^2ff_xg_xgh - 12x^4ff_xg_xgh^2 \\
\ \ - 12x^4f^2g_xh_xgh - 24x^2f^2g_{xx}gh
\Big] \alpha_* 
\\ \ \ \ \ \ \ \ \ \ \ \ \ \ \ \ \ \ \ \ \ \ \ \ \ \ \ \ \ \ \ \ + \\
\Big[
\ 4x^2\Lambda_*g^2 + 12fg^2 - x^4fg_x^2h + 2x^4fg_{xx}gh + 24x^2fg^2h + 6x^3f_xg^2h \\ 
\ \ + 6x^3fh_xg^2 + x^4f_xg_xgh + x^4fg_xxh_xg + 10x^3fg_xgh - 16g^2
\Big]
= 0
\end{array}
\label{eqn:EGB_fieldequation_psispi}
\end{equation}
\begin{equation}
\begin{array}{l}
\Big[
\ 16f^2g^2h - 4x^2f^2g_x^2h - 48x^2f^2g^2h^2 + 8xf^2h_xg^2 + 4x^4f^2g_x^2h^2 + 8xff_xg^2h \\
\ \ + 8x^2f^2g_{xx}gh + 24xf^2g_xgh - 8x^4f^2g_{xx}gh^2 - 64x^3ff_xg^2h^2 - 24x^3f^2h_xg^2h \\
\ \ + 4x^2f^2g_xh_xg + 4x^2ff_xg_xgh - 40x^4ff_xg_xgh^2 - 12x^4f^2g_xh_xgh - 8x^4f_{xx}fg^2h^2 \\
\ \ - 4x^5f_x^2g_xgh^2 + 2x^5f_xg_x^2fh^2 - 4x^5g_{xx}f_xfgh^2 - 4x^5f_{xx}g_xfgh^2 \\ 
\ \ - 12x^4f_xh_xfg^2h - 8x^4f_x^2g^2h^2 - 40x^3f^2g_xgh^2 - 6x^5f_xg_xh_xfgh
\Big] \alpha_* 
\\ \ \ \ \ \ \ \ \ \ \ \ \ \ \ \ \ \ \ \ \ \ \ \ \ \ \ \ \ \ \ \ + \\
\Big[
\ 4x^2\Lambda_xg^2 - 4fg^2 + 6x^3fh_xg^2 + 24x^2g^2h + 16x^3f_xh^2h - x^4fg_x^2h + 10x^3fg_xgh \\
\ \ + x^4f_xh_xg^2 + 2x^4f_xh_xgh + x^4fg_xh_xg + 2x^4fg_{xx}gh + 2x^4f_{xx}g^2h
\Big]
= 0
\end{array}
\label{eqn:EGB_fieldequation_thetatheta}
\end{equation}

\section{Series Solutions}
\label{sec:semianalytical_results}

We were unable to obtain an analytic  solution to the EGB field equations for the ansatz 
\eqref{eqn:EH5D_our_metric}. We therefore turn to  power-series expansions and numerical techniques to find  soliton solutions to  EGB gravity.

\subsection{Power-series solutions}

There are two particularly interesting power-series for the soliton, near infinity (the large-$r$ expansion) and
near $r=r_0$ (the near-$r_0$ expansion).  While not sufficient for demonstrating a full soliton solution these series are useful guides as to the behaviour of the solution in interesting physical limits.

\noindent For large-$r$ we find that
\begin{equation}
\begin{array}{llcl}
{} & f(r) &=& 1 + \frac{a_4}{r^4} - \frac{a_4 (4\alpha a_4 - b_4 \ell^2)}{2(4\alpha-\ell^2)r^8} - \frac{a_4 \ell^2(12\alpha a_4 - 48\alpha b_4 + 5 b_4\ell^2)}{15(4\alpha-\ell^2)r^{10}} + {\cal O}\left(\frac{1}{r^{10}}\right) \\ {} & g(r) &=& 1 + \frac{\ell^2}{r^2} + \frac{b_4}{r^4} - \frac{b_4 (4\alpha b_4 - \ell^2 a_4)}{2(4 \alpha - \ell^2)r^8} + \frac{a_4 \ell^2 (252\alpha a_4 - 5 b_4 \ell^2 + 48 \alpha b_4)}{45(4\alpha - \ell^2)r^{10}} + {\cal O}\left(\frac{1}{r^{10}}\right) \\ \
{} & h(r) &=& 1 + \frac{\ell^2}{r^2} + \frac{b_4}{r^4} - \frac{b_4 (-7\ell^2 a_4 + 12 \alpha b_4)}{6(4 \alpha - \ell^2)r^8} + \frac{a_4 \ell^2 (756\alpha a_4 - 5 b_4 \ell^2 + 336 \alpha b_4)}{45(4\alpha - \ell^2)r^{10}} + {\cal O}\left(\frac{1}{r^{10}}\right) \\  \end{array}
\label{eqn:EGB_EH_fgh_pow_large_r}
\end{equation}
where $4\alpha-\ell^2 \neq 0$ ($4\alpha_* - 1 \neq 0$).

There are two free parameters $(a_4,b_4)$ in the solution, governing the falloff rate of the metric functions.  
Their falloff rates suggest that the mass of the soliton, $\mathbb{M}$, is governed by these two quantities.  
Using the conformal formalism based on the electric Weyl tensor of \cite{Ashtekar2000-ConservedQuantities,Das2000-MoreConservedQuantities} to compute the conserved quantities, the conformal mass in Einstein gravity is given by
\begin{equation}
\mathbb{M} = -\frac{\pi(3b_4-a_4)}{8\ell^2\text{G}p}
\label{eqn:mass_soliton_einstein}
\end{equation}
Setting $a_4=-r_0^4$ and $b_4=0$ in \eqref{eqn:mass_soliton_einstein}, we recover the soliton mass from the counterterm subtraction method \cite{Clarkson2006-5DSoliton} and from the Hamiltonian formalism \cite{Copsey2007-Bubbles2}.

To obtain the near-$r_0$ power-series solution, we impose the conditions  $f(r_0)=0$, $g(r_0) \neq 0$, and $h(r_0) \neq 0$ 
at the bubble-edge $r_0$.  We then find 
\begin{equation}
\begin{array}{lcrl}
f(r) &=& {}      &A_1 (r-r_0) + A_2 (r-r_0)^2 + A_3 (r-r_0)^3 + \mathcal{O}\left((r-r_0)^4\right) \\ \\
g(r) &=& B_0 \ \ + &B_1 (r-r_0) + B_2 (r-r_0)^2 + B_3 (r-r_0)^3 + \mathcal{O}\left((r-r_0)^4\right) \\ \\
h(r) &=& C_0 \ \ + &C_1 (r-r_0) + C_2 (r-r_0)^2 + C_3 (r-r_0)^3 + \mathcal{O}\left((r-r_0)^4\right)
\end{array}
\label{eqn:EGB_EH_fgh_pow_near_r0_ansatz}
\end{equation}
where $\Lambda = -\frac{6}{\ell^2} + \frac{12\alpha}{\ell^4}$   and $B_0 \neq 0$ and $C_0 \neq 0$.
 The coefficients in the preceding expression are extremely cumbersome and we relegate them to Appendix \ref{app:large_r_pow_series_coeff}.

Before moving to a numerical solution of the equations, we revisit the issue of metric regularity for the EGB case.
The Kretschmann scalar $\mathcal{K} = R^{abcd}R_{abcd}$,  
\begin{eqnarray}
\mathcal{K} &=&
\frac{1}{4 \ell^4 g r^4}
\big( 
4g^4r^8f_{rr}^2h^2-4r^8gh^2f^2g_{rr}g_r^2+
32g^4r^7f_rh^2f_{rr}+
4r^8g^2h^2fg_{rr}g_rf_r \nonumber \\ && +
4r^8g^2hf^2g_{rr}g_rh_r+
16g^4r^6h^2ff_{rr}+
16r^6g^4h_r^2f^2+r^8h^2f^2g_r^4+88r^6g^4f_r^2h^2 \nonumber \\ && +
160r^4g^4h^2f^2+16r^6g^3h^2f^2g_{rr}+
8r^7g^3h^2fg_{rr}f_r+
4r^8g^2h^2f^2g_{rr}^2 \nonumber \\ && +
24r^7g^2h^2f^2g_rg_{rr}+
8r^7g^3f_r^2h^2g_r+4r^7g^3h_r^2f^2g_r+
8r^7g^3hf^2g_{rr}h_r \nonumber \\ && +
8g^4r^7f_{rr}hh_rf+
r^8g^2g_r^2h_r^2f^2+
16g^4r^7f_r^2hh_r+
48f_r^2h\ell^2g^4r^4-
12r^7gh^2f^2g_r^3 \nonumber \\ && +
8r^7g^3h_rfg_rf_rh+
2r^8g^2g_r^2h_rff_rh+
4g^4r^8f_{rr}hh_rf_r+
4g^4r^7h_r^2ff_r \\ && -
128g^4\ell^2r^2fh+
32g^4r^2f^2h\ell^2+
2r^8g^2g_r^2f_r^2h^2+
40r^6g^2h^2f^2g_r^2+
64r^5g^4hf^2h_r \nonumber \\ && +
160r^5g^4h^2ff_r+
96r^5g^3h^2f^2g_r+
176g^4\ell^4f^2-
384g^4f\ell^4+
g^4r^8h_r^2f_r^2 \nonumber \\ &&+
256\ell^4g^4 -
2r^8hf^2g_r^3gh_r+
8r^7g^2hf^2g_r^2h_r+
12r^7g^2h^2fg_r^2f_r \nonumber \\ &&+
64r^6g^4h_rff_rh +
48r^6g^3h^2fg_rf_r-
32g^4r^3f_rhf\ell^2-2r^8h^2fg_r^3gf_r \nonumber \\ && +
32r^6g^3hf^2g_rh_r 
\big) \nonumber
\label{eqn:kretschmann_scalar_general}
\end{eqnarray}
and will be finite for all $r \geq r_0$ provided $g(r) \neq 0$ for all $r > r_0$. The only remaining possible singularities will be conical singularities at $r = r_0$ and string singularities.  The former will not be present provided $\psi$ has a period of $2\pi/\mathcal{P}$ where
\begin{equation}
\mathcal{P}^2 \equiv \frac{r_0^4 A_1^2 R^2 C_0}{16 \ell^2}
\label{eqn:special_P}
\end{equation}
String-singularities will be eliminated  at the north and south poles if  $\psi$ has a period of $4\pi/p$ where $p \in \mathbb{Z} \setminus \{0\}$.  Together these conditions imply
\begin{equation}
A_1 = \sqrt{\frac{4 \ell^2 p^2}{r_0^4 C_0}}
\label{eqn:metric_reg_condition}
\end{equation}

\noindent This is a generalization of the regularity condition for the Einsteinian case \cite{Clarkson2006-OddSoliton}.  For $\alpha=0$ it
is straightforward to show that   $A_1 = \frac{4}{r_0}$ and $C_0 = 1+\frac{\ell^2}{r_0^2}$ giving
\begin{equation}
r_0^2 = \ell^2\left(\frac{p^2}{4}-1\right)
\label{eqn:metric_reg_condition_einstein}
\end{equation}
which is the regularity condition (\ref{EHmatch}), with $a=r_0$.

\section{Numerical Results}
\label{sec:numerical_results}

In this section we present the numerical solutions of the EGB field equations.
The power-series solutions from Section \ref{sec:semianalytical_results} provide a useful guide for approximating the initial conditions (hereafter IC) of the large-$r$ and near-$r_0$ field equations.
Both positive and negative values of the cosmological constant are possible, depending on the choice of Gauss-Bonnet coefficient $\alpha_*$.

We have compared the full numerical solutions with their respective power-series expansion as a cross-check on our numerical work. As shown more fully in Appendix \ref{app:powerseries_expansion}, the large-$r$ numerical solution agrees very well with the corresponding power series solution when expanded to order $1/r^{10}$, even for reasonably small $r$. The near-$r_0$ power-series agrees well with the numerical solution when the corresponding power series solution is expanded up to $(r-r_0)^3$ order, but can quickly deviate from the numerical solution when $r $ is appreciably larger than $r_0$. We use the power series expansions given in Section \ref{sec:semianalytical_results} for the initial conditions. 

With consistent results from the power-series approximations, we now present the full numerical solutions below.

\subsection{Initial Conditions and Numerical Procedure}
\label{subsec:initial_conditions}

\noindent We find that there is numerical instability if the values of $|\alpha_*|$ are too large. 
From the series solutions \eqref{eqn:EGB_EH_fgh_pow_large_r} we see that $\alpha_* \neq 1/4$, so we shall consider only small $\alpha_* < 1/4$ henceforth. 
We can choose freely the 
parameter   $p$, which is a positive integer from condition \eqref{eqn:metric_reg_condition} required to satisfy
 metric regularity conditions.  
 In Einstein gravity, we have the bound $p \ge 3$ from the regularity condition \eqref{eqn:metric_reg_condition_einstein}. The analogous lower bound on $p$ in EGB gravity follows from condition \eqref{eqn:metric_reg_condition} since the coefficients $A_1$ and $C_0$ will depend on $r_0$ and $\ell$, which we do not know a priori. For sufficiently small $\alpha_*$ we expect that $p=5$ will be sufficient, and we shall make
 this choice throughout the paper.  It is a straightforward (albeit tedious) exercise to numerically obtain solutions for larger $p$. 

  All other parameters   are constrained by the shooting method, a trial-and-error method of tuning the values of chosen IC parameters that are not known a priori, so that when the equations are numerically integrated
away from the starting point, the solution will satisfy desired asymptotic properties.   The IC parameters $\{a^*_4,b^*_4\}$ and $\{B_0,C_0\}$ will be used to tune the required desired properties of the solution for the large-$r$ and near-$r_0$ solutions, respectively.   For numerical integration beginning near-$r_0$,  we use the shooting method to require $\{f,g,h\}$ to follow the required large-$r$ asymptotic conditions.  Conversely, for 
large-$r$ numerical integration, the starting point is at $r=\infty$, and the shooting method is used to set the value of the bubble edge  at $r=r_0$. 

For convenience, we list here the various parameters we employ for both large-$r$ and near-$r_0$ numerical integration of the field equations.
\begin{itemize}[]
\item {\bf \ $\alpha_*$ } \\
This is the dimensionless Gauss-Bonnet coefficient, chosen to be small to avoid issues of  
numerical instability as noted above.
\item {\bf \ $p$ } \\ 
This is a non-zero integer that emerges from the the metric regularity condition \eqref{eqn:metric_reg_condition}, whose choice selects one of a  countably infinite set of  EH soliton solutions in EGB gravity.
The 5D Einsteinian EH soliton required that $p \ge 3$ \cite{Clarkson2006-5DSoliton}.
For practical purposes, we will use $p=5$ for our numerical results.
\item \ $B_0$ \\
This the value of $g(r=r_0)$; it  must be positive and  converge to $1 + \ell^2 / r_0^2$ in the $\alpha \rightarrow 0$ limit.
The near-$r_0$ shooting method will constrain this parameter.
\item \ $C_0$ \\
Similar to $B_0$, this is exactly the value of $h(r=r_0)$ and should be positive and converging to $1 + \ell^2 / r_0^2$ in the $\alpha \rightarrow 0$ limit.
The near-$r_0$ shooting method will constrain this parameter.
\item \ $x_0$ \\
This is the dimensionless radius of the EH soliton bubble edge $r_0$ in units of $\ell$ defined by $x_0 \equiv r_0/\ell$.
The spacetime   exists only for $x \ge x_0$ (equivalently $r \ge r_0$).
The large-$r$ shooting method will constrain this parameter.
\item {\bf \ $a_4$ } \\
This parameter contributes to the mass of the soliton.
In the Einsteinian limit $\alpha \to 0$,   $a_4 = - r_0^4$, where $r_0$ is the soliton radius.  This interpretation
will not necessarily hold for nonzero $\alpha$ and $b_4$. The large-$r$ shooting method will constrain this parameter, whose rescaled version  is $a^*_4 = a_4 / \ell^4$.
\item {\bf \ $b_4$ } \\
The $b_4$ parameter also contributes to the mass of the soliton; in Einstein gravity the relevant expression is
that of Eqn \eqref{eqn:mass_soliton_einstein}.
We will consider both $b_4 = 0$ and $b_4 \neq 0$ cases. The large-$r$ shooting method will constrain this parameter, whose rescaled version  is $b^*_4 = b_4 / \ell^4$.
\item \ $s_0$ \\
This is the reciprocal dimensionless radius of the EH soliton bubble edge, defined by $s_0 = \ell/r_0$.
To numerically analyze the  $r \rightarrow \infty$ limit, we make a variable change $r \mapsto s = \ell/r_0$ and analyze its $s \rightarrow 0$ behavior. Note that the spacetime is valid for $s \le s_0$ only.  
\end{itemize}

\noindent The outline of the numerical procedure is as follows. For some $\alpha_*$ small, we choose $p=5$. We then make an initial choice of the IC parameters $\{a^*_4,b^*_4\}$ (large-$r$) or $\{B_0,C_0\}$ (near-$r_0$). The other 
parameters in the respective power-series solutions   are then fully determined from this guess.
We then apply  the Runge-Kutta-Fehlberg method to integrate the field equations ${\mathcal{E}_{rr},\mathcal{E}_{\psi\psi},\mathcal{E}_{\theta\theta}}$,  with the starting point at $r=r_0$ (near-$r_0$) or at $r=\infty$ (large-$r$). 
If the required asymptotic behaviour of the soliton (when starting at $r=r_0$) or bubble edge behaviour (when starting at $r=\infty$) is satisfied, we are done and have obtained the numerical solution.
If they are not satisfied within a given tolerance (we use a tolerance of $\epsilon = 10^{-4}$), then we repeat the process with improved guesses of the ICs and calculate the numerical solution.

\subsection{Large-$r$ behavior}
We employ the parameter $s \equiv \ell/r < 1$ to investigate the large-$r$ behavior of our solutions.
Our choice of $p=5$ implies that  $s^{\alpha=0}_0 = 1/x^{0}_0 = \left(p^2/4 -1\right)^{-1/2} \mathop{\approx}_{p=5} 0.4364$ in the Einsteinian case. 
In Einstein gravity, we also have $a_4^{*,\alpha=0} = -x_0^4 \mathop{=}_{p=5} -27.5625$ and $b_4^{*,\alpha=0} = 0$.
We find that the field equations for large-$r$ are numerically more difficult than near-$r_0$, so we only display results that we numerically trust.

The free parameters chosen for our large-$r$ numerical results are given in Table 1.  We present our results for
the parameter set $\mu \in \{a^*_4,b^*_4,s_0\}$ in terms of their deviations $\Delta \mu = \mu - \mu^{\alpha=0}$ relative their values in Einstein gravity.
\begin{table}[H]
\centering
\begin{tabular}{ C{0.5cm} C{2.0cm} | C{0.9cm} | C{2.4cm} | C{2.4cm} | C{2.4cm} | C{2.4cm} |  }
\cline{3-7}
{} & {} & \multicolumn{5}{|c|}{\textbf{Numerical Parameters -- large-$r$}} \\
\cline{3-7}
{} & {} & \multicolumn{1}{|c|}{$\alpha_*=0$} & \multicolumn{2}{|c|}{$\alpha_*=-0.04$} & \multicolumn{2}{|c|}{$\alpha_*=0.002$} \\
\hline
\multicolumn{2}{|c|}{\bf \texttt{Figure Number}} & {\bf \texttt{\#\ref{fig:fgh_sol_large_r_num_alpha0_1}}} & {\bf \texttt{\#\ref{fig:fgh_sol_large_r_num_alpha-001_1}}} & {\bf \texttt{\#\ref{fig:fgh_sol_large_r_num_alpha-001_2}}} & {\bf \texttt{\#\ref{fig:fgh_sol_large_r_num_alpha0001_1}}} & {\bf \texttt{\#\ref{fig:fgh_sol_large_r_num_alpha0001_2}}} \\
\hline
\multicolumn{1}{|c|}{$p$} & Free & $5$ & $5$ & $5$ & $5$ & $5$ \\
\hline
\multicolumn{1}{|c|}{$\Delta a_4^*$} & Shooting method & $0$ & $7.80$ & $7.80$ & $-0.20$ & $-6.30$ \\
\hline
\multicolumn{1}{|c|}{$\Delta b_4^*$} & Shooting method & $0$ & $0$ & $-2.96$ & $0$ & $0.200$ \\
\hline
\multicolumn{1}{|c|}{$\Delta s_0$} & Calculated & $0$ & $5.17 \times 10^{-2}$ & $2.00 \times 10^{-2}$ & $2.20 \times 10^{-3}$ & $-2.11 \times 10^{-2}$ \\
\hline
\multicolumn{1}{|c|}{$\Delta \mathbb{M}^*$} & Calculated & $ 0$ & $1.56$ & ${3.34}$ & ${-0.04}$ & ${-1.38}$ \\
\hline
\end{tabular}
\caption{   Numerical values  for the large-$r$ numerical solution. The symbols used are $s_0 = 1/x_0$, $s^{\alpha=0}_0 \approx 0.436436$, $a_4^{*,\alpha=0} = -27.5625$, $b_4^{*,\alpha=0} = 0$, and $\mathbb{M}^{*,\alpha=0} = -5.51$. Deviations from the $\alpha=0$ values are defined by $\Delta a_4^* = a_4^* - a_4^{*,\alpha=0}$, $\Delta b_4^* = b_4^* - b_4^{*,\alpha=0}$, and $\Delta s_0 = s_0 - s_0^{\alpha=0}$. The mass deviation parameter $\Delta \mathbb{M}^* = \mathbb{M}^* - \mathbb{M}^{*,\alpha=0}$ is in units of $\pi \ell^2/8G$, where the latter term  is given in  Eqn \eqref{eqn:mass_soliton_einstein}.} 
\label{tab:numerical_values_large_r}
\end{table}

\noindent We find that the numerical integration is  much more sensitive to the choice of $b^*_4$ than $a^*_4$, and so require more fine-tuning of $b_4^*$.

\begin{figure}[H]
\vspace{-10pt}
\begin{center}
 \includegraphics[scale=0.45]{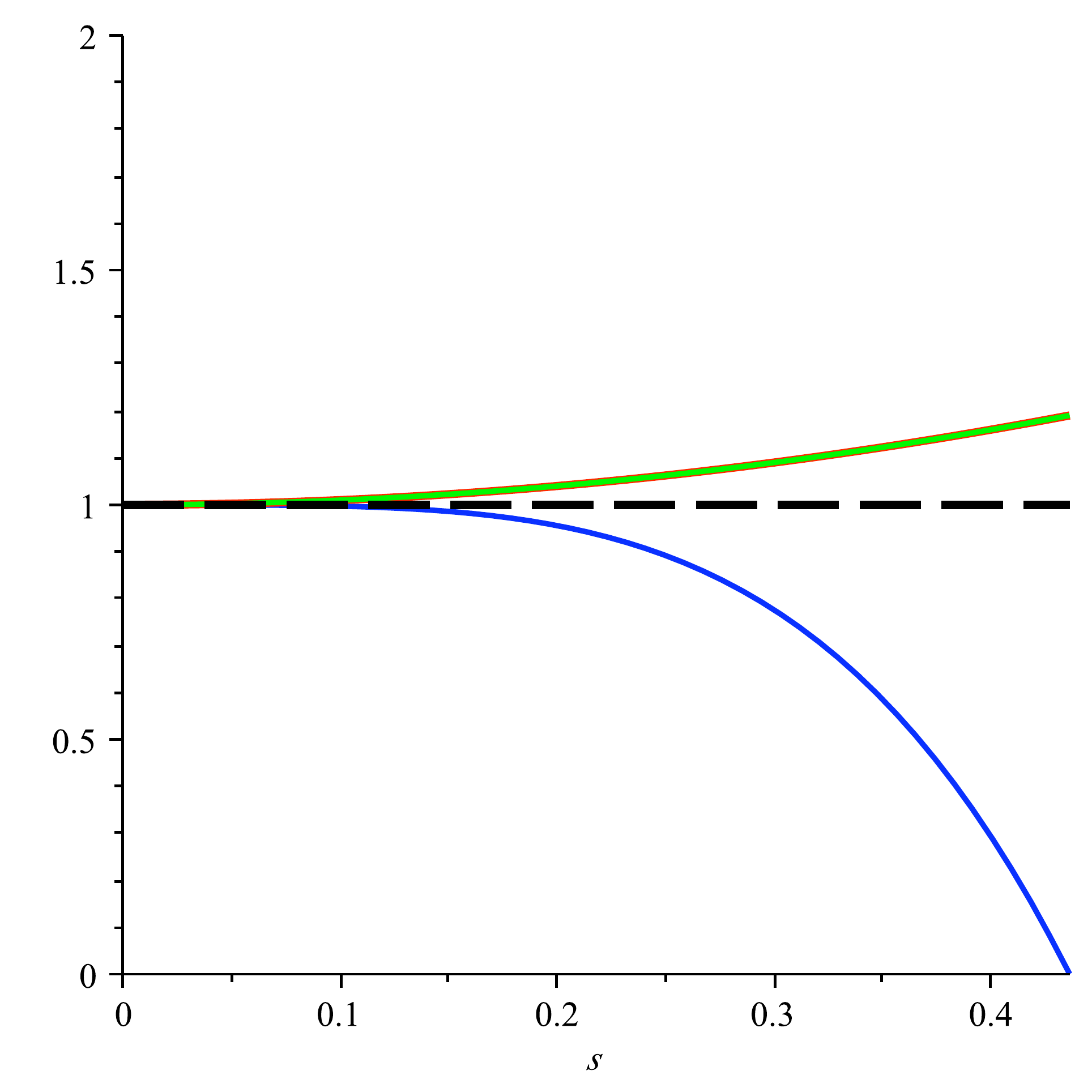}
\end{center}
\vspace{-20pt}
\caption{Large-$r$ numerical solutions $\{f,g,h\}$ for $\alpha_* = 0$ where $s=\frac{\ell}{r}$. We have $a_4^* = a_4^{*,\alpha=0}$, $b^*_4 = 0$ and $s_0 = s_0^{\alpha=0}$ here. The color coding is $f(s) = \text{blue}$, $g(s) = \text{red}$, and $h(s) = \text{green}$.}
\vspace{-10pt}
\label{fig:fgh_sol_large_r_num_alpha0_1}
\end{figure}

\begin{figure}[H]
\vspace{-10pt}
\begin{center}
 \includegraphics[scale=0.45]{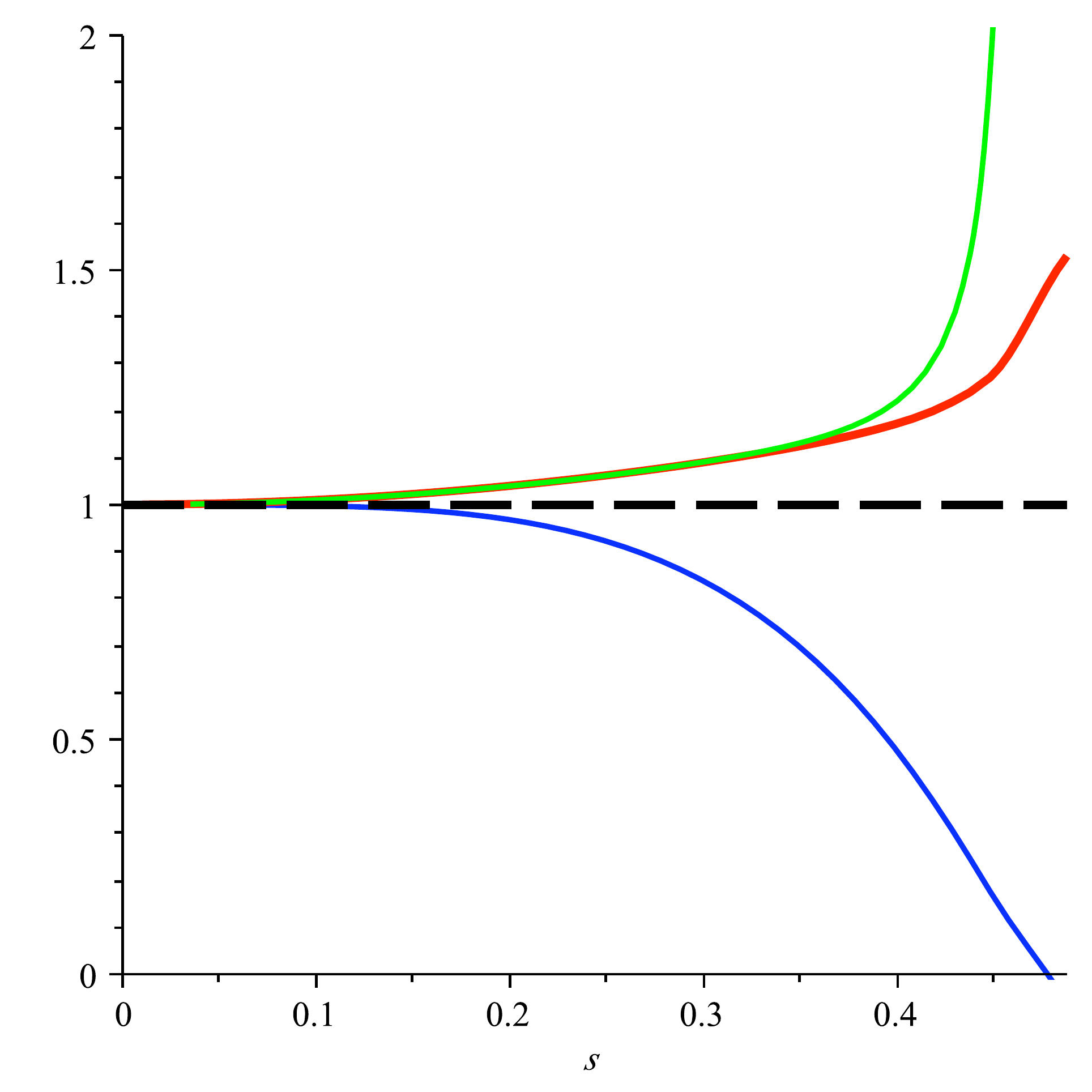}
\end{center}
\vspace{-20pt}
\caption{Large-$r$ numerical solutions $\{f,g,h\}$ for $\alpha_* = -0.04$ where $s=\frac{\ell}{r}$. We have $a_4^* = a_4^{*,\alpha=0} + 7.80$, $b^*_4 = 0$, $s_0 = s_0^{\alpha=0} + 0.0517$ here. The color coding is $f(s) = \text{blue}$, $g(s) = \text{red}$, and $h(s) = \text{green}$.}
\vspace{-10pt}
\label{fig:fgh_sol_large_r_num_alpha-001_1}
\end{figure}

\begin{figure}[H]
\vspace{-10pt}
\begin{center}
 \includegraphics[scale=0.45]{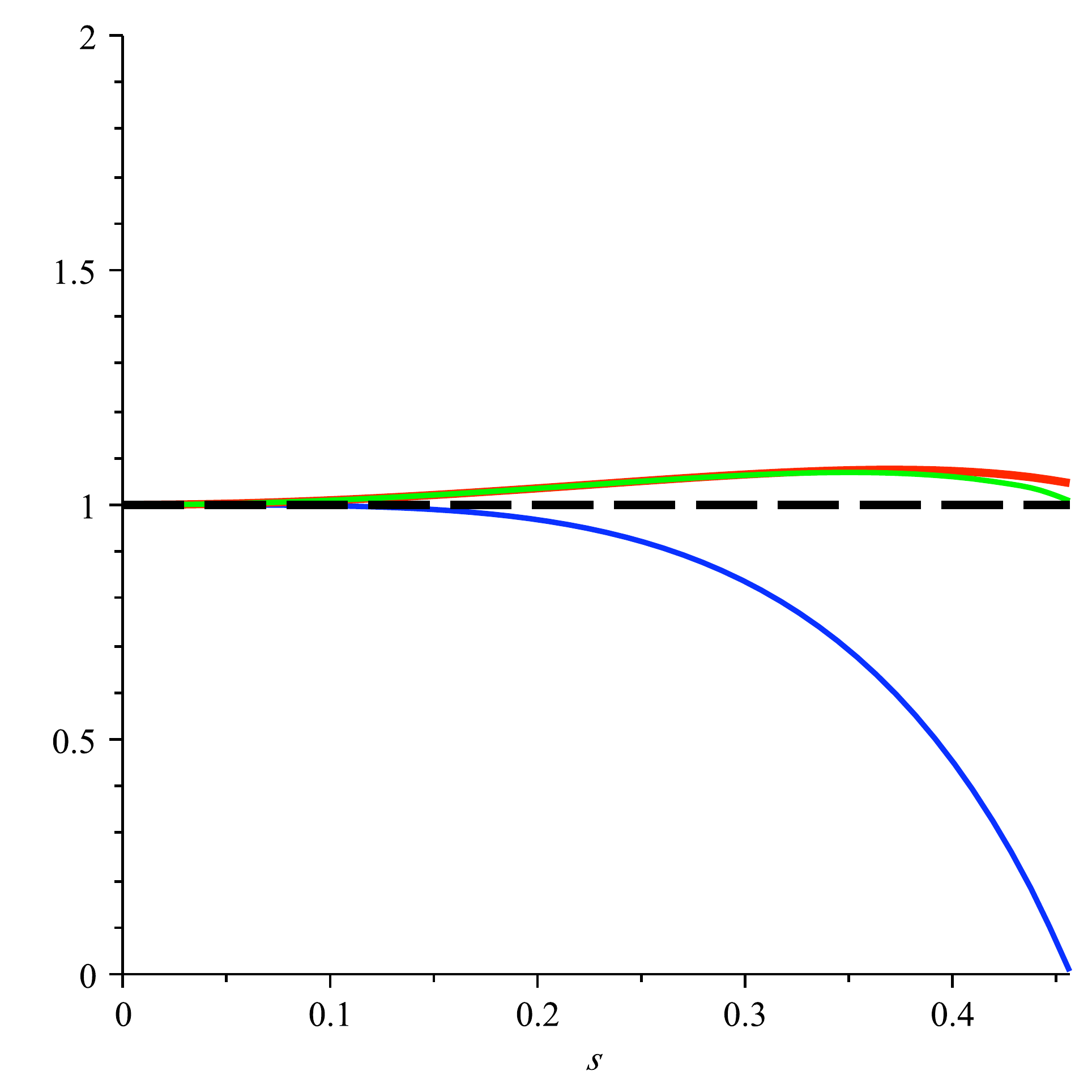}
\end{center}
\vspace{-20pt}
\caption{Large-$r$ numerical solutions $\{f,g,h\}$ for $\alpha_* = -0.04$ where $s=\frac{\ell}{r}$. We have $a_4^* = a_4^{*,\alpha=0} + 7.80$, $b^*_4 = -2.96$, $s_0 = s_0^{\alpha=0} + 0.02$ here. The color coding is $f(s) = \text{blue}$, $g(s) = \text{red}$, and $h(s) = \text{green}$.}
\vspace{-10pt}
\label{fig:fgh_sol_large_r_num_alpha-001_2}
\end{figure}

\begin{figure}[H]
\begin{center}
\vspace{-10pt}
 \includegraphics[scale=0.45]{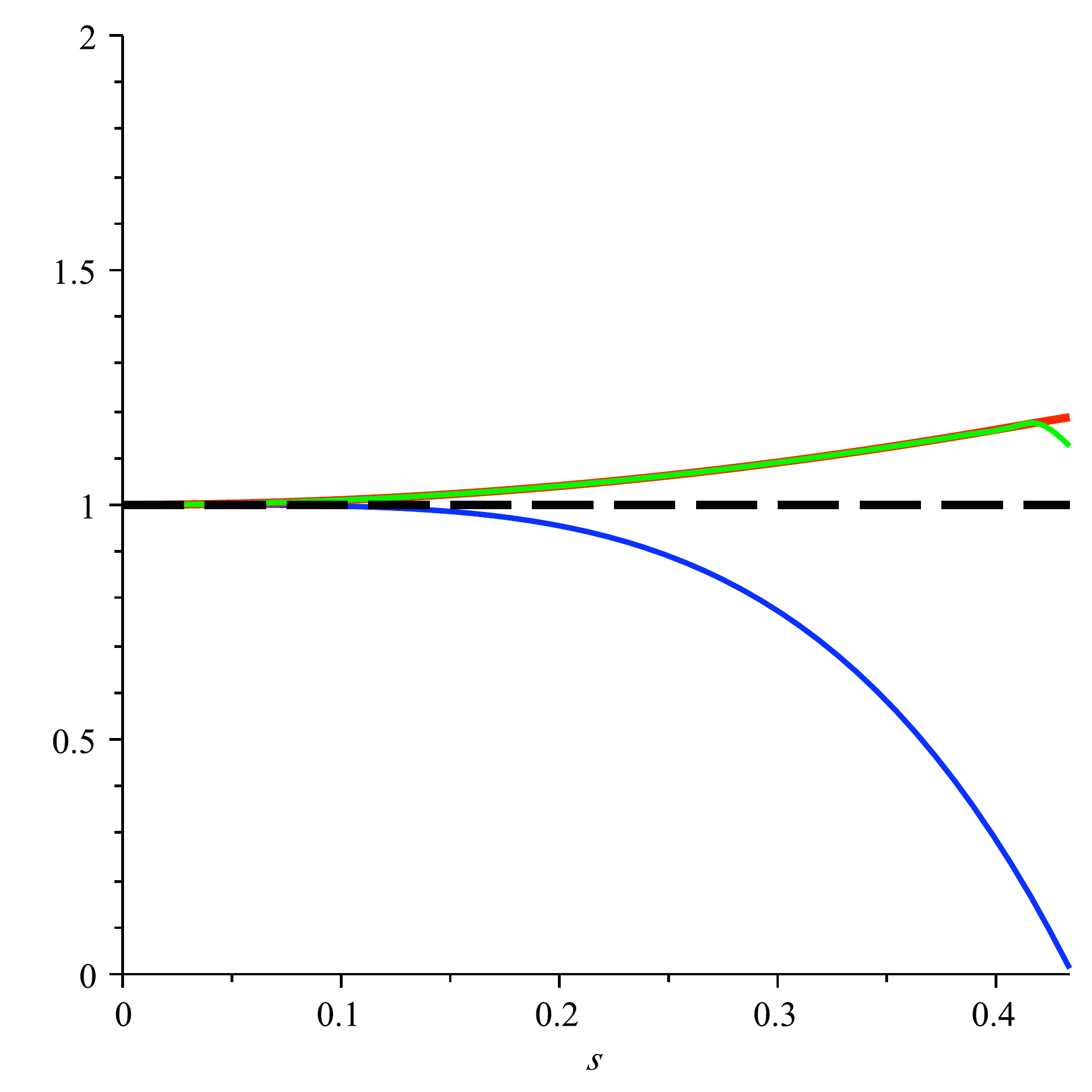}
\end{center}
\vspace{-20pt}
\caption{Large-$r$ numerical solutions $\{f,g,h\}$ for $\alpha_* = 0.002$ where $s=\frac{\ell}{r}$. We have $a_4^* = a_4^{*,\alpha=0} - 0.20$, $b^*_4 = 0$, $s_0 = s_0^{\alpha=0} - 0.0022$ here. The color coding is $f(s) = \text{blue}$, $g(s) = \text{red}$, and $h(s) = \text{green}$.}
\vspace{-10pt}
\label{fig:fgh_sol_large_r_num_alpha0001_1}
\end{figure}

\begin{figure}[H]
\begin{center}
\vspace{-10pt}
 \includegraphics[scale=0.45]{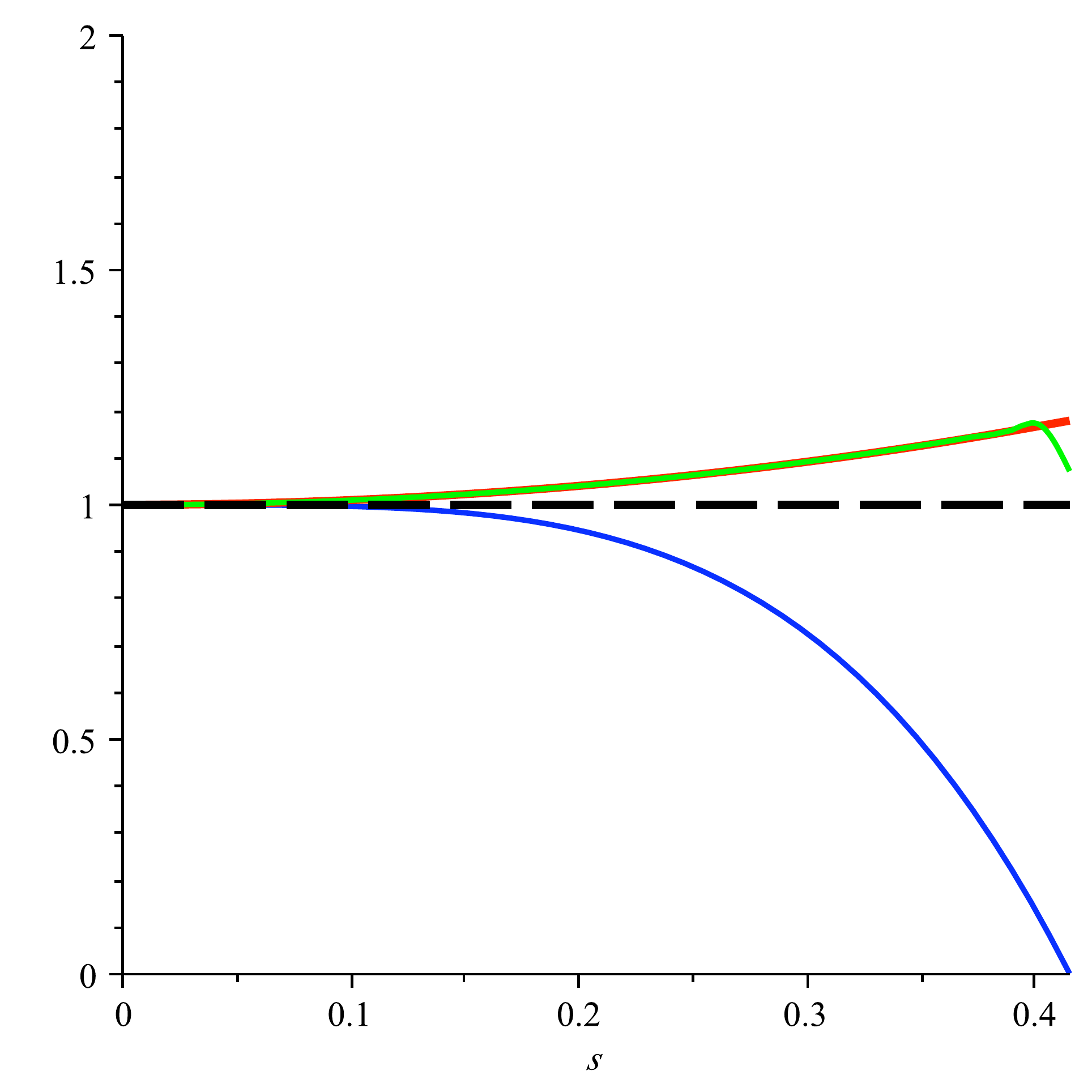}
\end{center}
\vspace{-20pt}
\caption{Large-$r$ numerical solutions $\{f,g,h\}$ for $\alpha_* = 0.002$ where $s=\frac{\ell}{r}$. We have $a_4^* = a_4^{*,\alpha=0} - 6.30$, $b^*_4 = 0.20$, $s_0 = s_0^{\alpha=0} - 0.0211$ here. The color coding is $f(s) = \text{blue}$, $g(s) = \text{red}$, and $h(s) = \text{green}$.}
\vspace{-10pt}
\label{fig:fgh_sol_large_r_num_alpha0001_2}
\end{figure}

Figure \ref{fig:fgh_sol_large_r_num_alpha0_1} -- Figure \ref{fig:fgh_sol_large_r_num_alpha0001_2} depict the solutions $\{f(s),g(s),h(s)\}$ where $s=\ell/r$ for $\alpha_* = 0,-0.04,0.002$ from Table \ref{tab:numerical_values_large_r}, respectively. 
We see that all metric functions approach $1$ in the $s \rightarrow 0$ limit (at $r \rightarrow \infty$) as required.  
Our solutions are valid all the way from infinity to the edge of the soliton, whose values are given
in Table \ref{tab:numerical_values_large_r} for the various cases.

 We also see from Fig. \ref{fig:fgh_sol_large_r_num_alpha-001_2} that a nonzero value for $b_4$ can have a dramatic effect relative to the $b_4=0, \alpha=-0.04$ solutions shown in Fig. \ref{fig:fgh_sol_large_r_num_alpha-001_1}. This effect is less pronounced for the $\alpha=0.002$   solutions,  as shown in Fig. \ref{fig:fgh_sol_large_r_num_alpha0001_1} and Fig. \ref{fig:fgh_sol_large_r_num_alpha0001_2} possibly due to the decreasing magnitude of $\alpha$. 

\subsection{Near-$r_0$ behavior}

The free and calculated parameters with their numerical values for the near-$r_0$ solution are shown in Table \ref{tab:numerical_values_near_r0}.
In the Einsteinian case,  $B_0^{\alpha=0} = C_0^{\alpha=0} = 1 + (1/x^{\alpha=0}_0)^2 = 1 + \left( p^2/4-1 \right)^{-1} \mathop{\approx}_{p=5} 1.1905$ where $x^{\alpha=0}_0 = \left(p^2/4-1\right)^{1/2}\mathop{\approx}_{p=5} 2.2913$.
As in the large-$r$ case, we express our results in terms of deviations from Einstein gravity.
\begin{table}[H]
\centering
\begin{tabular}{ C{1.0cm} C{3.5cm} | C{1.6cm} | C{2.7cm} | C{2.7cm} |  }
\cline{3-5}
{} & {} & \multicolumn{3}{|c|}{\textbf{Numerical Parameters -- near-$r_0$}} \\
\cline{3-5}
{} & {} & $\alpha_*=0$ & $\alpha_*=-0.04$ & $\alpha_*=+0.002 $ \\
\hline 
\multicolumn{2}{|c|}{\bf \texttt{Figure Number} } & {\bf \texttt{\#\ref{fig:fgh_sol_near_r0_num_alpha0_1}}} & {\bf \texttt{\#\ref{fig:fgh_sol_near_r0_num_alpha-004_1}}} & \bf{ \texttt{\#\ref{fig:fgh_sol_near_r0_num_alpha0002_1}}} \\
\hline
\multicolumn{1}{|c|}{$p$} & Free Parameter & $5$ & $5$ & $5$ \\
\hline
\multicolumn{1}{|c|}{$\Delta B_0$} & Shooting method & $0$ & $-0.1423$ & $0.0087$ \\
\hline
\multicolumn{1}{|c|}{$\Delta C_0$} & Shooting method & $0$ & $-0.1782$ & $0.0065$ \\
\hline
\multicolumn{1}{|c|}{$\Delta x_0$} & Calculated & $0$ & $-0.2442$ & $0.0116$ \\
\hline
\end{tabular}
\caption{  Numerical values  for the near-$r_0$ solution, where  $B^{\alpha=0}_0 = C^{\alpha=0}_0 \approx 1.1905$ and $x^{\alpha=0}_0 = 2.2913$. Deviations from the $\alpha=0$ values are defined by $\Delta B_0 = B_0 - B_0^{\alpha=0}$, $\Delta C_0 = C_0 - C_0^{\alpha=0}$, and $\Delta x_0 = x_0 - x_0^{\alpha=0}$.   }
\label{tab:numerical_values_near_r0}
\end{table}
 The only free parameter in the near-$r_0$ solution is $p$; all other parameters can be obtained numerically via the shooting method and from solving the constraint equations.

\begin{figure}[H]
\vspace{-10pt}
\begin{center}
 \includegraphics[scale=0.45]{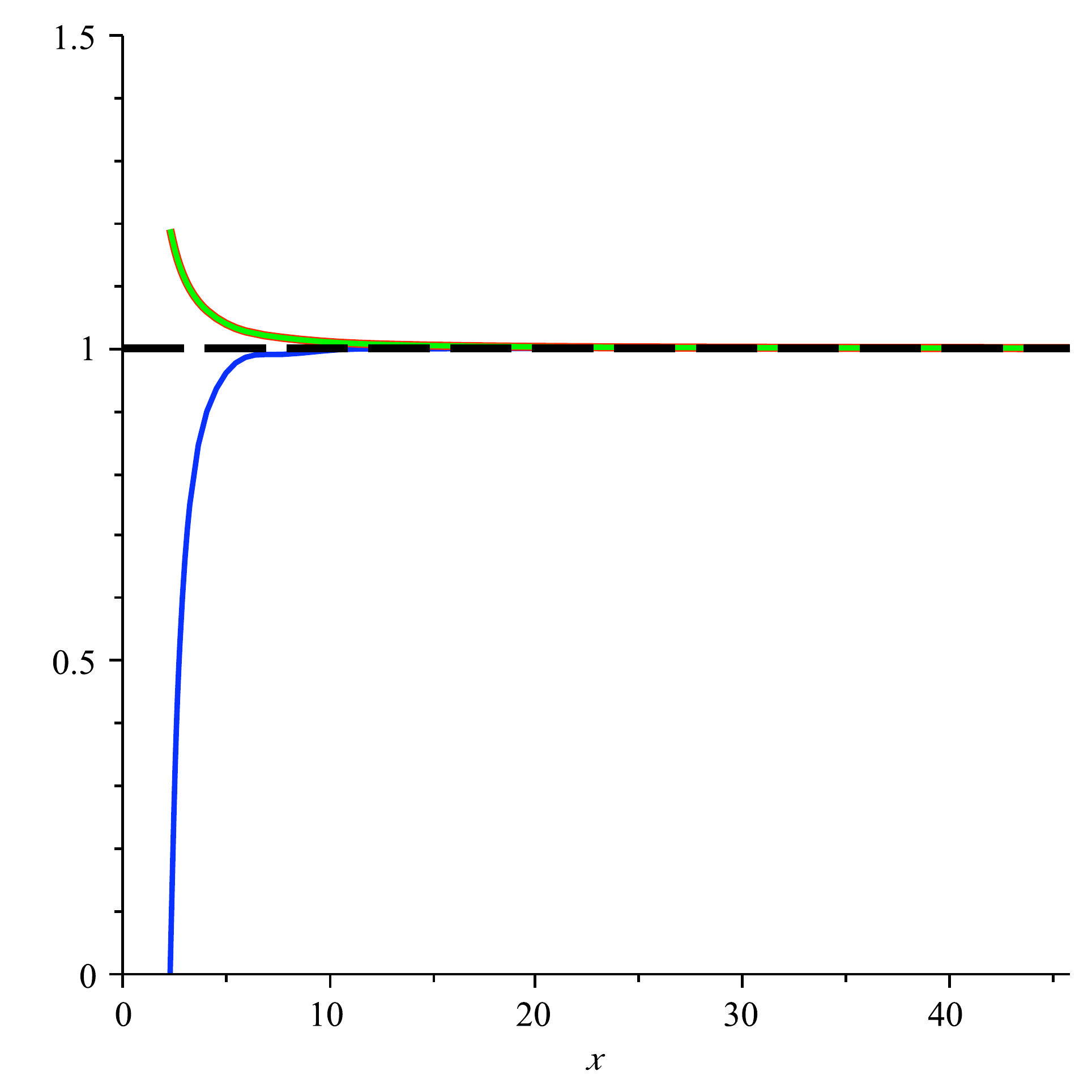}
\end{center}
\vspace{-20pt}
\caption{Near-$r_0$ numerical solutions of $\{f,g,h\}$ for $\alpha_* = 0$, $p=5$, $B_0 = B_0^{\alpha=0}$, $C_0 = C_0^{\alpha=0}$, and $x_0 = x_0^{\alpha=0}$. The color coding is $f(x) = \text{blue}$, $g(x) = \text{red}$, and $h(x) = \text{green}$.}
\vspace{-10pt}
\label{fig:fgh_sol_near_r0_num_alpha0_1}
\end{figure}

\begin{figure}[H]
\vspace{-10pt}
\begin{center}
 \includegraphics[scale=0.45]{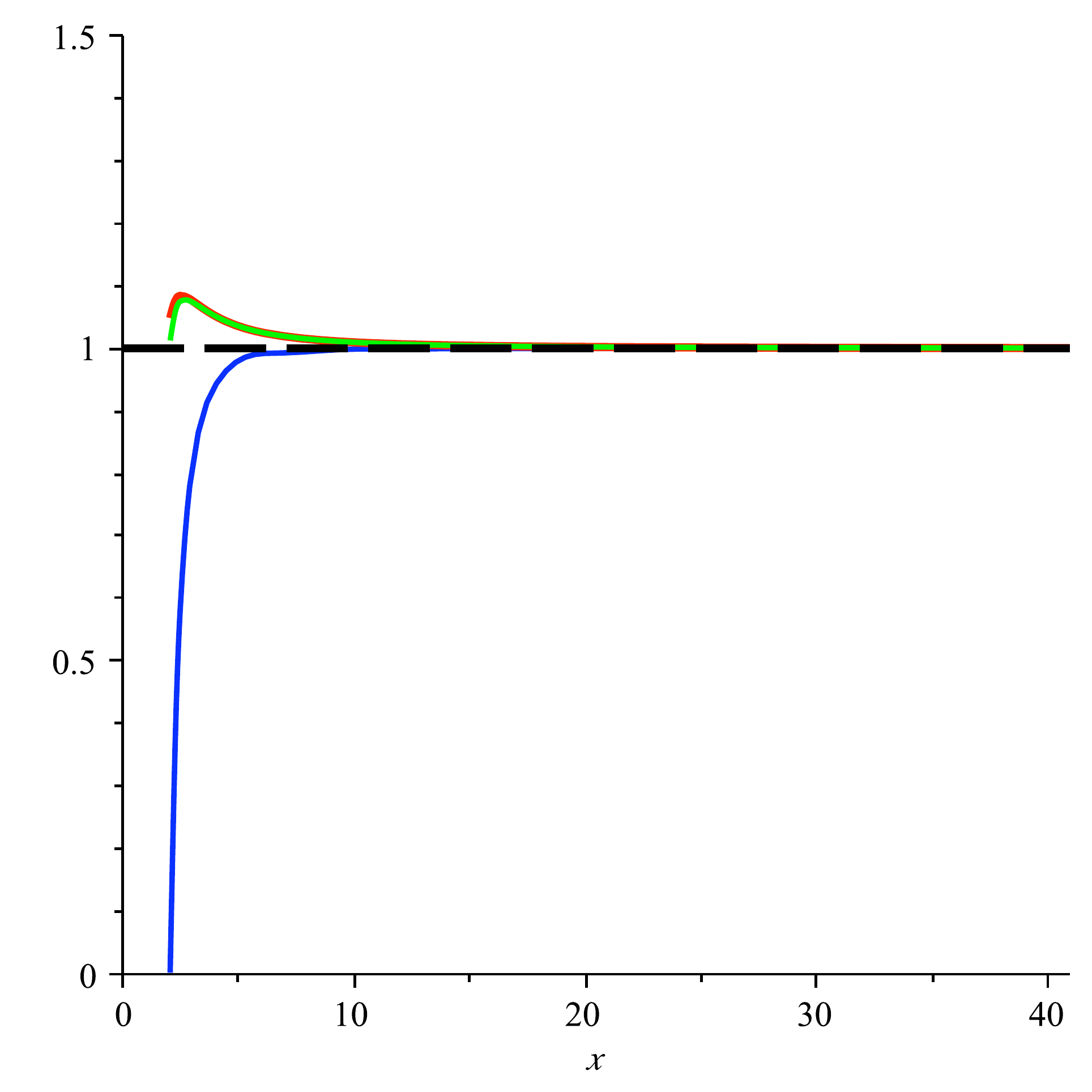}
\end{center}
\vspace{-20pt}
\caption{Near-$r_0$ numerical solutions $\{f,g,h\}$ for $\alpha_* = -0.04$, $p=5$, $B_0 = B_0^{\alpha=0} - 0.1423$, $C_0 = C_0^{\alpha=0} - 0.1782$, and $x_0 = x_0^{\alpha=0} - 0.2442$. The color coding is $f(x) = \text{blue}$, $g(x) = \text{red}$, and $h(x) = \text{green}$.}
\vspace{-10pt}
\label{fig:fgh_sol_near_r0_num_alpha-004_1}
\end{figure}

\begin{figure}[H]
\vspace{-10pt}
\begin{center}
 \includegraphics[scale=0.45]{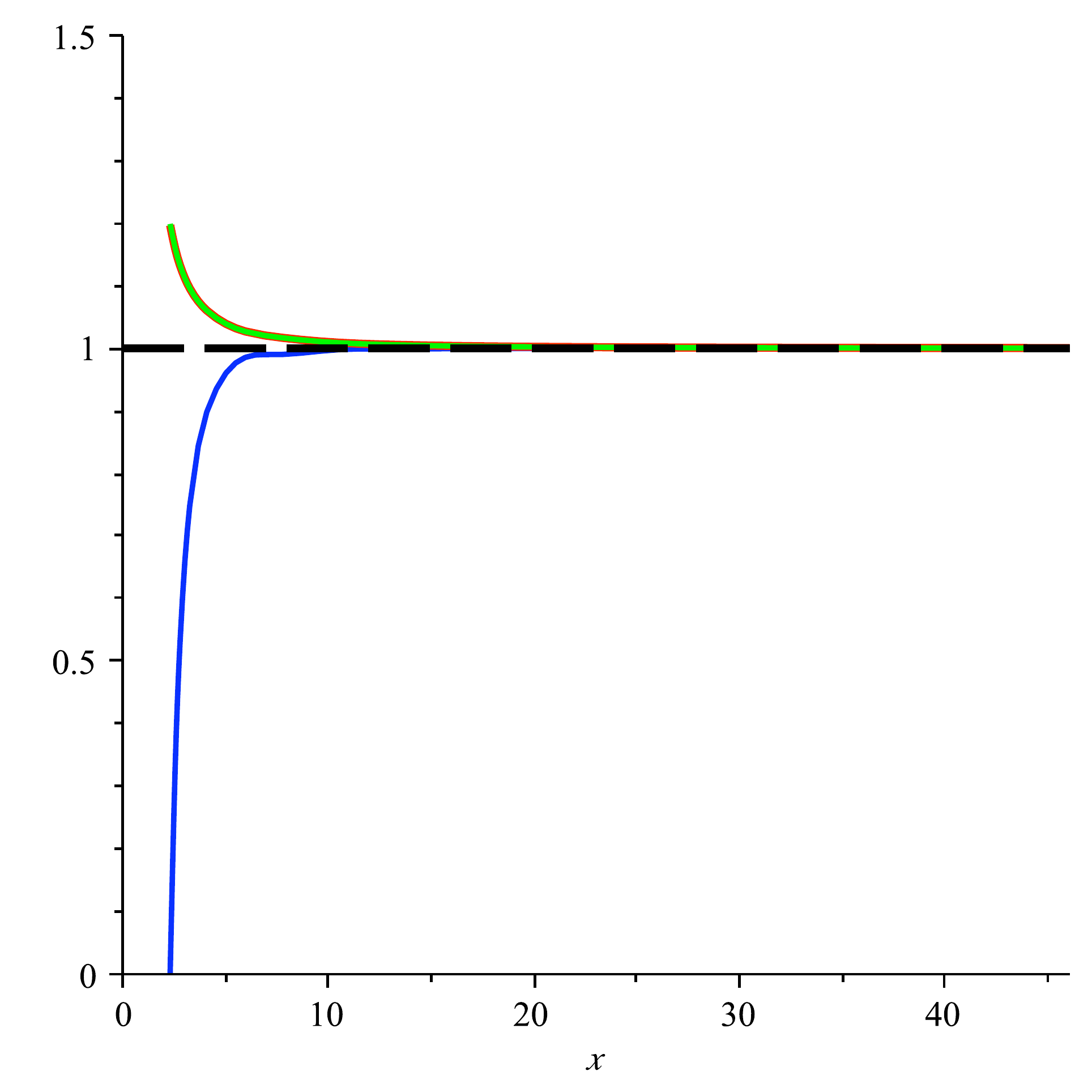}
\end{center}
\vspace{-20pt}
\caption{Near-$r_0$ numerical solutions $\{f,g,h\}$ for $\alpha_* = +0.002$, $p=5$, $B_0 = B_0^{\alpha=0} + 0.0087$, $C_0 = C_0^{\alpha=0} + 0.0065$, and $x_0 = x_0^{\alpha=0} + 0.0116$. The color coding is $f(x) = \text{blue}$, $g(x) = \text{red}$, and $h(x) = \text{green}$.}
\vspace{-10pt}
\label{fig:fgh_sol_near_r0_num_alpha0002_1}
\end{figure}

\section{Conclusions}
\label{sec:conclusions_and_discussions}

We have semi-analytically and numerically found   five-dimensional Eguchi-Hanson soliton solutions in Einstein-Gauss-Bonnet (EGB) gravity.   We have illustrated the typical form of the metric functions for small positive and negative values of $\alpha$, integrating from both large-$r$ to the edge of the soliton, and from the edge of the soliton
to infinity.   These numerical solutions are fully consistent with large-$r$ and the near-$r_0$  power-series solutions with  
2 free parameters $\alpha$ and $p$.

We have found numerical evidence (in the context of the 
large-$r$ problem) that $b_4$ can be non-vanishing in EGB gravity.  This indicates a broader class of soliton solutions, in which both $a_4$ and $b_4$ can be nonzero.  A full exploration of these solutions, along with
a study of a broad range of $\alpha_*$  (going beyond the choice we believe falls in a safe range of stability), remain interesting subjects for future investigation.

\bigskip \noindent \textbf{Acknowledgements} \\
We would like to acknowledge helpful discussions with K. Copsey and D. Kubiz\v{n}\'{a}k.
This work was supported in part by the Natural Sciences and Engineering Research Council of Canada.

\section{Appendix}
\appendix

\section{Numerical Solutions Compared to Power Series Expansions}
\label{app:powerseries_expansion}

We present here a comparison of our numerical solutions to the power-series expansion of our results, up to $1/r^{10}$ for the large-$r$ solution and up to $(r-r_0)^3$ in the near-$r_0$ solution for $\alpha=0$. 

The large-r power-series is very effective when expanded up to $1/r^{10}$, but the near-$r_0$ solutions fail to demonstrate proper asymptotic behavior when expanded only to order $(r-r_0)^3$.

\begin{figure}[H]
\begin{center}
\subfigure[The large-$r$ power-series for $\alpha_*=0$. ]{
 \includegraphics[scale=0.4]{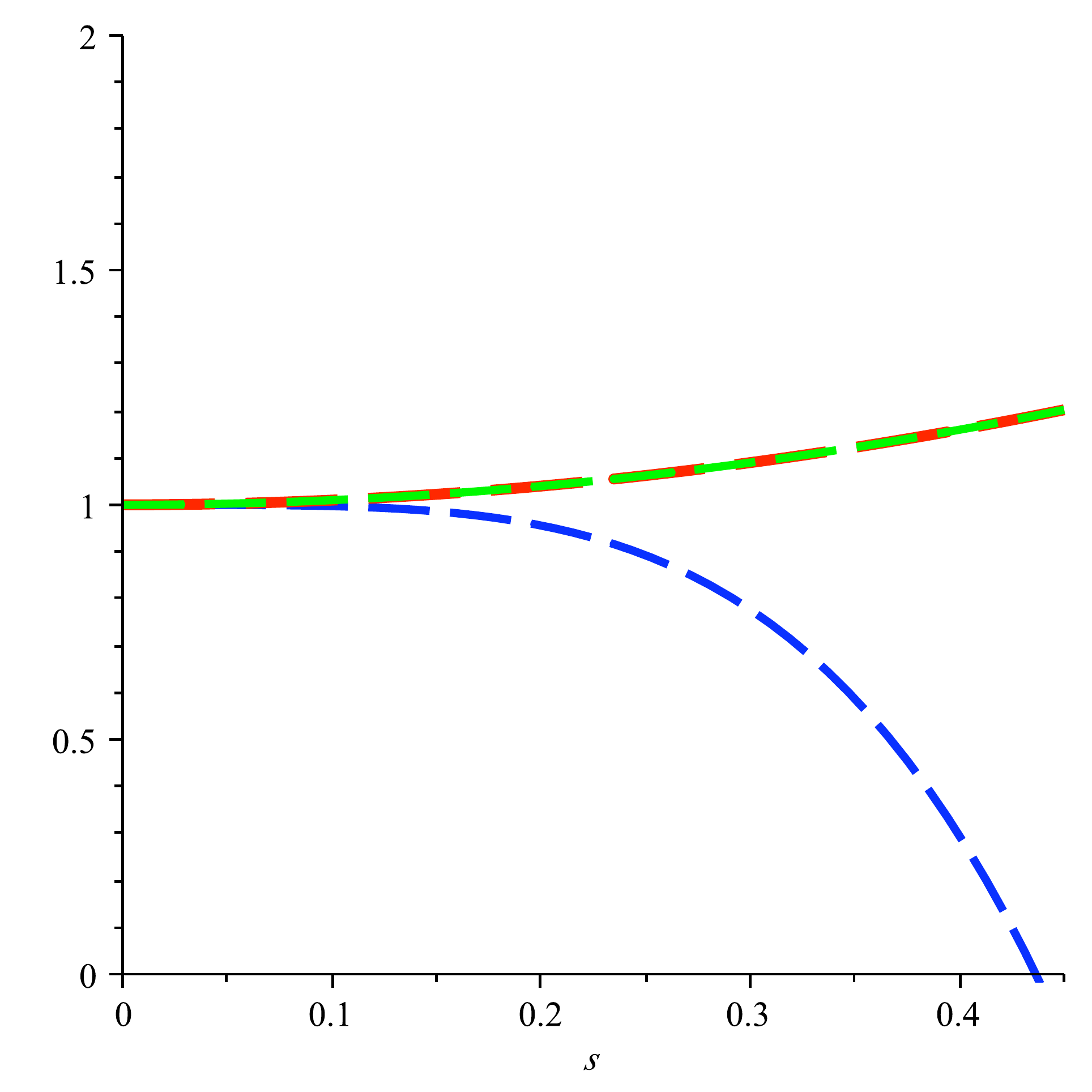}
 \label{subfig:larger_pow_num1}
}
\subfigure[The large-$r$ numerical solution for $\alpha_*=0$. ]{
 \includegraphics[scale=0.4]{larger_alpha0__1.pdf}
 \label{subfig:larger_alpha0}
}
\subfigure[The large-$r$ power-series for $\alpha_*=-0.04$. ]{
 \includegraphics[scale=0.4]{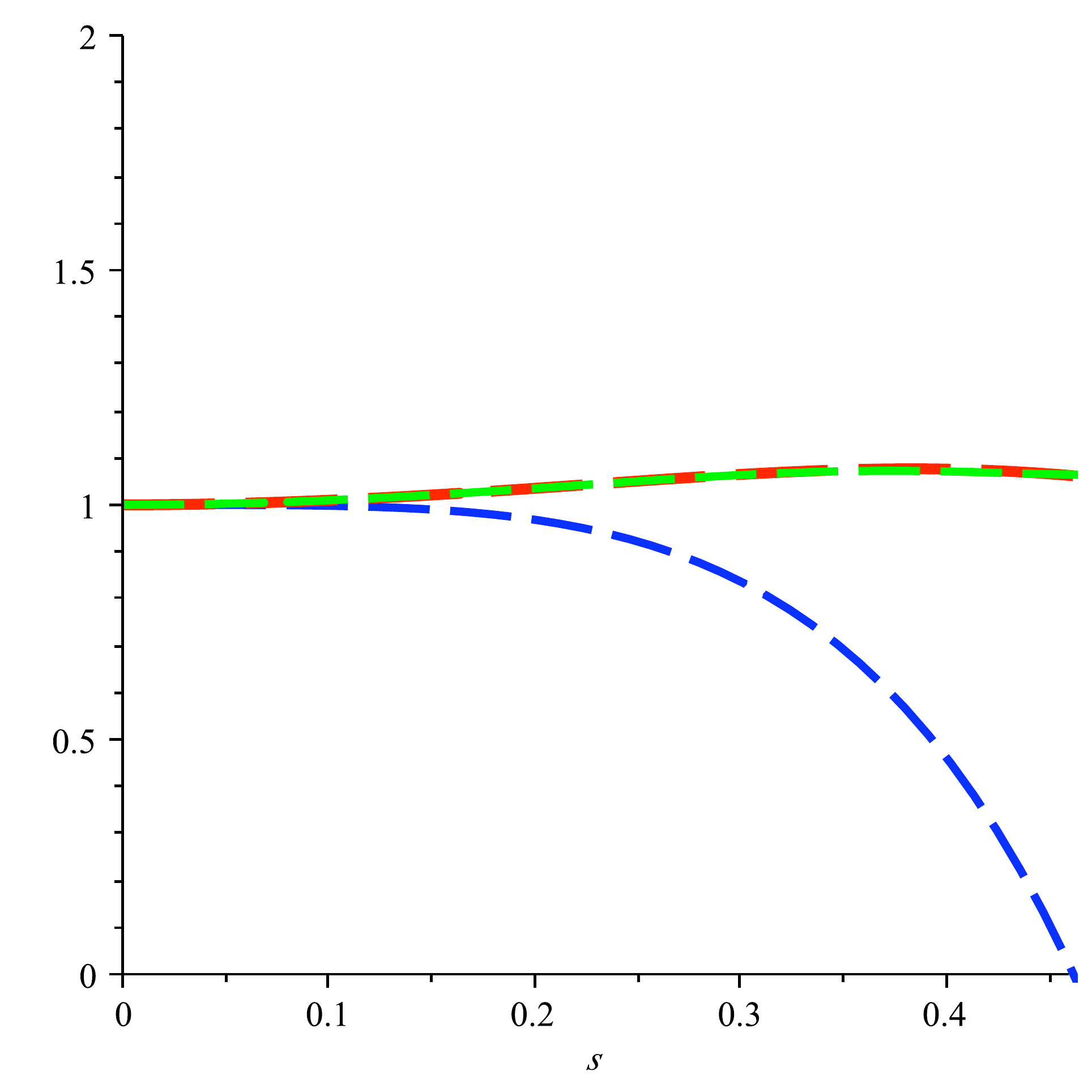}
 \label{subfig:larger_pow_num1_alpha004}
}
\subfigure[The large-$r$ numerical solution for $\alpha_*=-0.04$. ]{
 \includegraphics[scale=0.4]{larger_alphan004_b4__3.pdf}
 \label{subfig:larger_alpha004}
}
\end{center}
\caption{ The large-r solution comparison of the power-series in expanded up to $1/r^{10}$ (dashed lines) and full numerical solution (solid lines) for $\alpha_*=0$ and $\alpha_*=-0.04$ with a colour coding of $f=$ blue, $g=$ red, and $h=$ green.}
\end{figure}

\begin{figure}[H]
\begin{center}
\subfigure[The near-$r_0$ power-series for $\alpha_*=0$. ]{
  \includegraphics[scale=0.4]{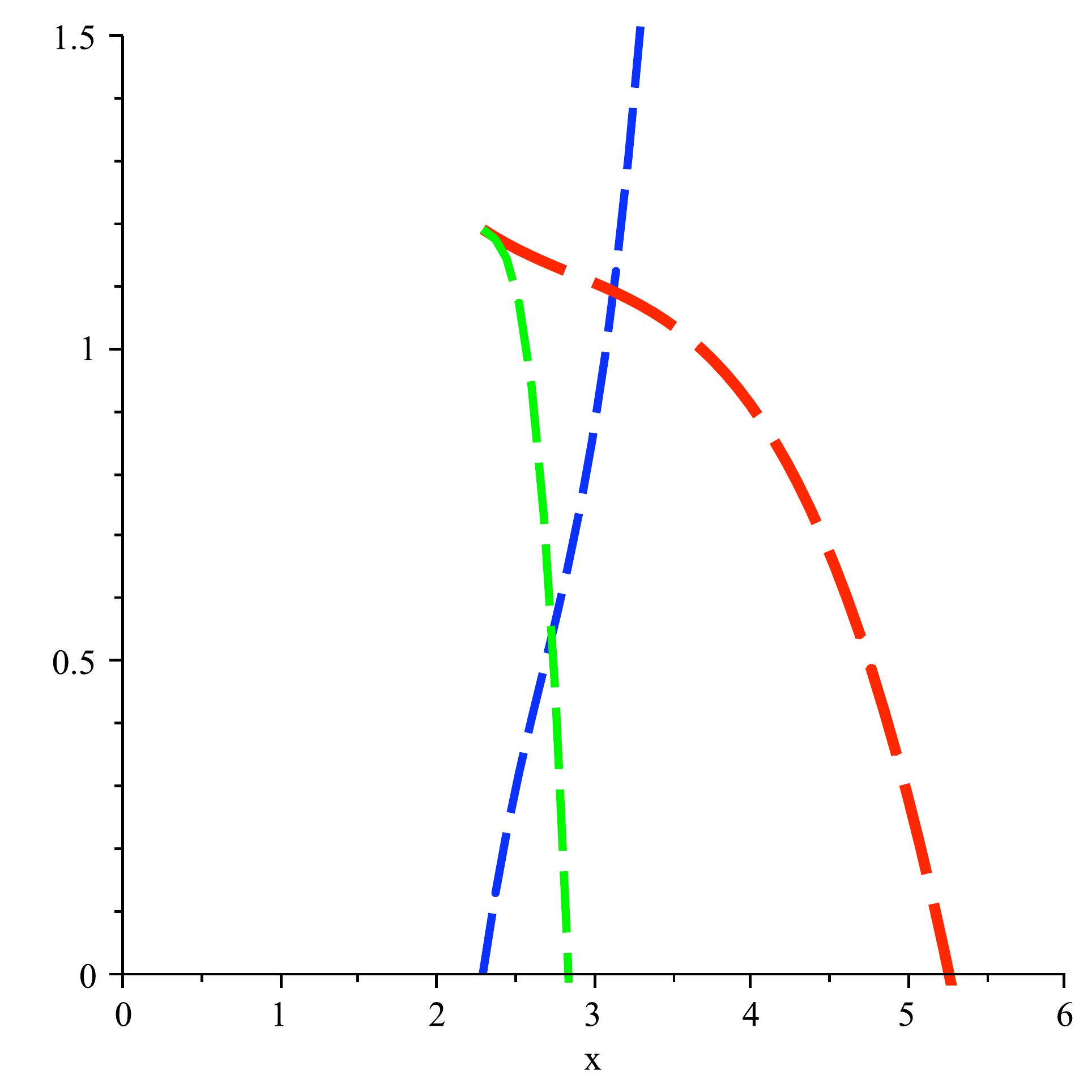}
 \label{subfig:nearr0_alpha0_pw}
}
\subfigure[The near-$r_0$ numerical solution for $\alpha_*=0$. ]{
 \includegraphics[scale=0.4]{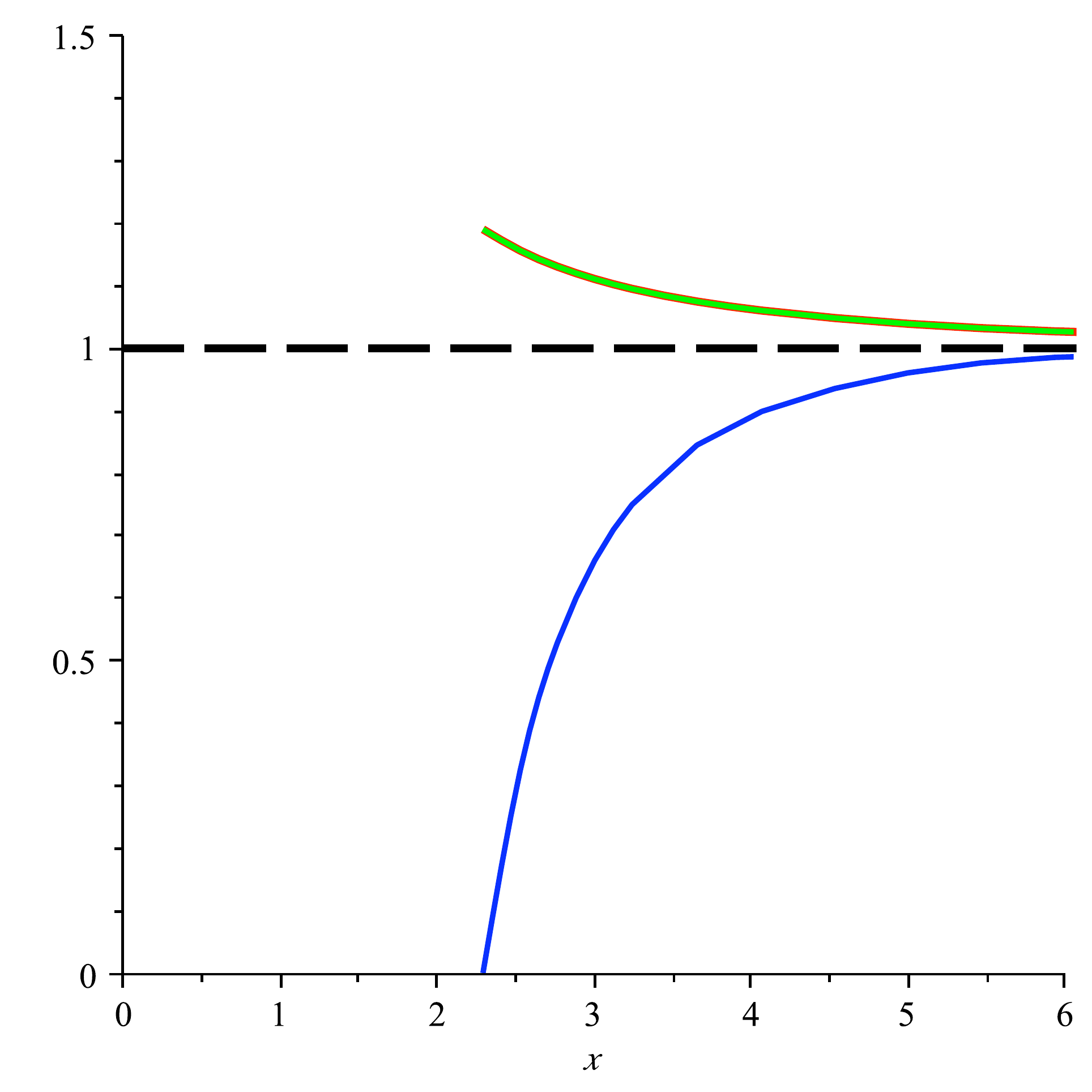}
 \label{subfig:nearr0_alpha0_num}
}
\subfigure[The near-$r_0$ power-series for $\alpha_*=-0.04$. ]{
  \includegraphics[scale=0.4]{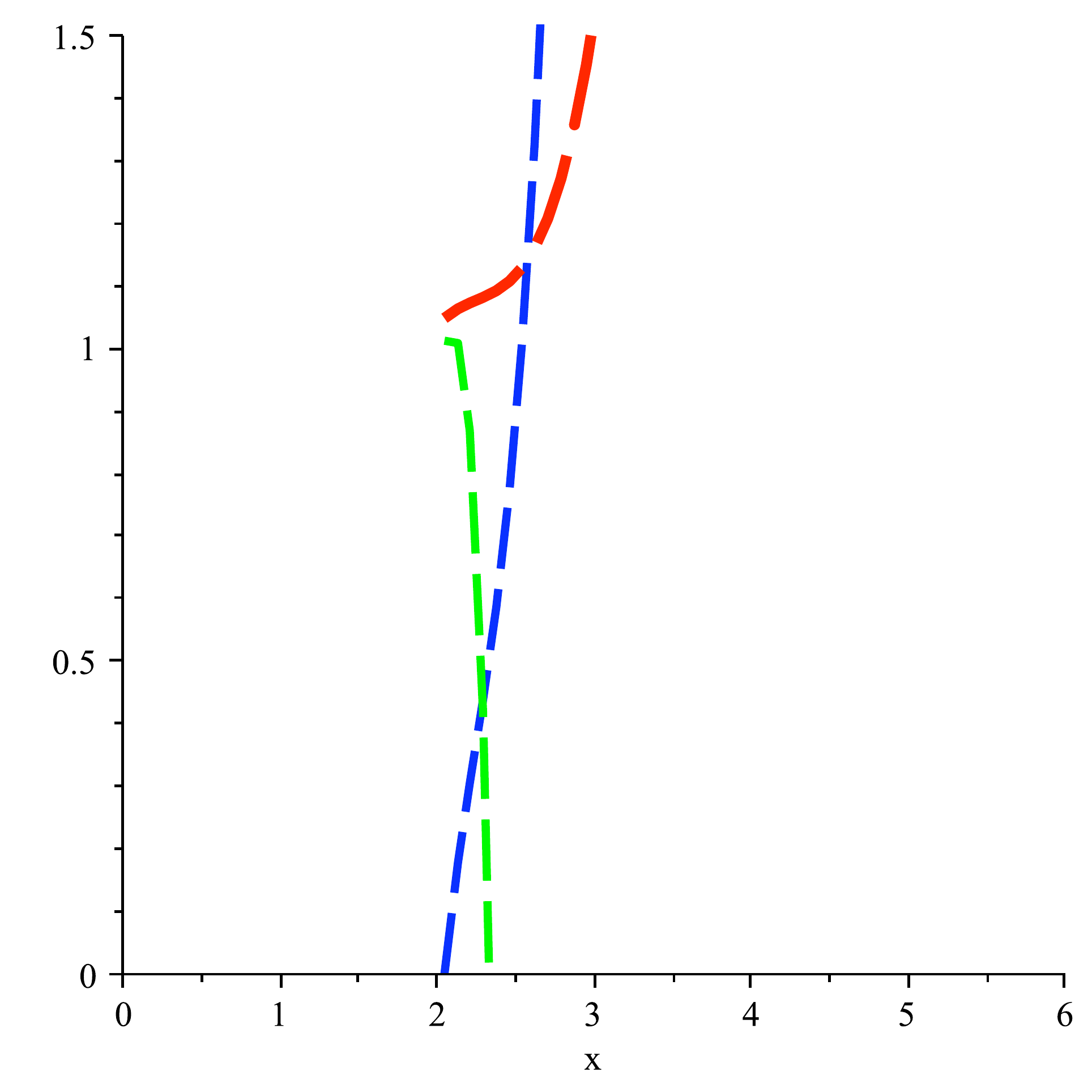}
 \label{subfig:nearr0_alpha004_pw}
}
\subfigure[The near-$r_0$ numerical solution for $\alpha_*=-0.04$. ]{
 \includegraphics[scale=0.4]{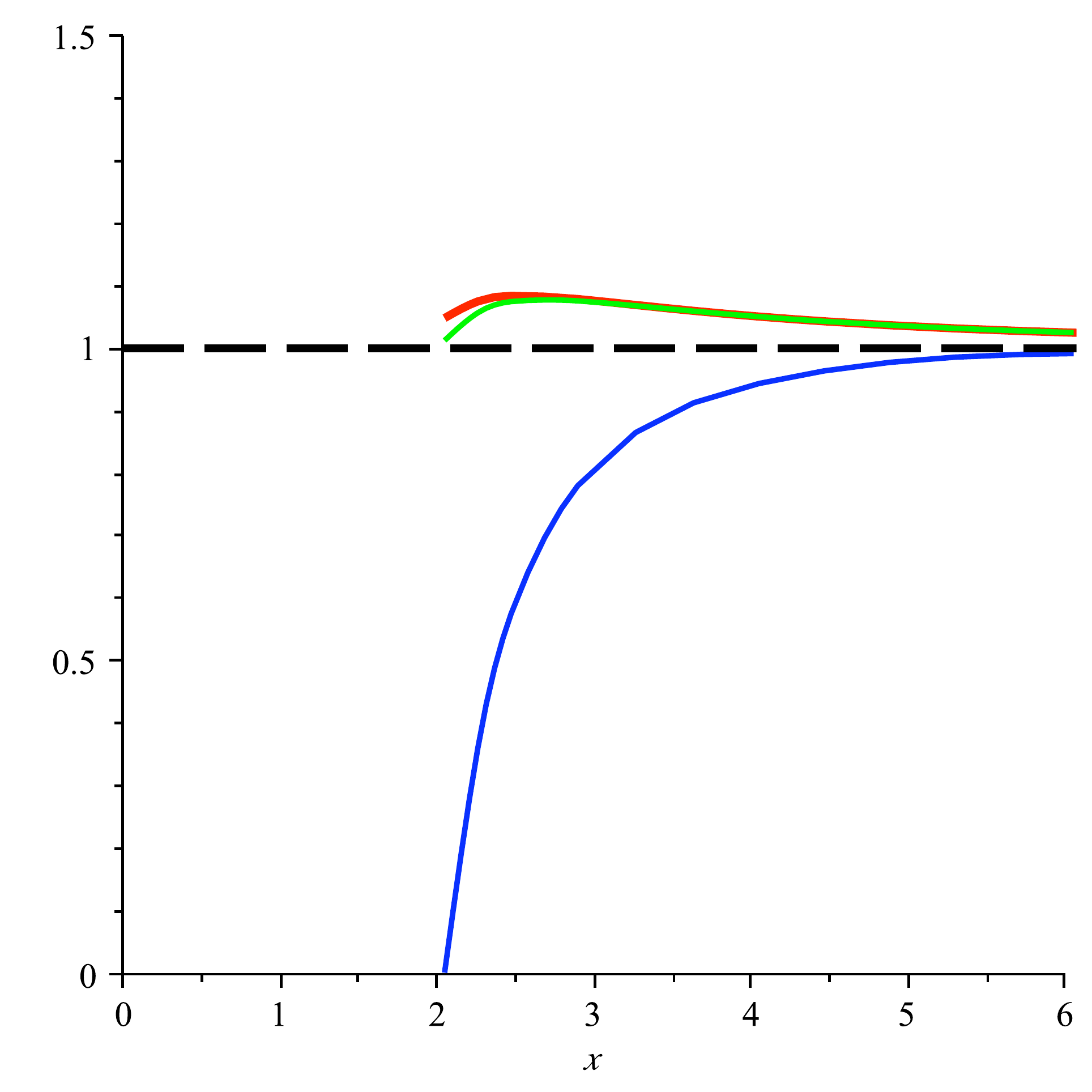}
 \label{subfig:nearr0_alpha004_num}
}
\end{center}
\caption{ The near-$r_0$ solution comparison of the power-series in expanded up to $(r-r_0)^3$ (dashed lines) and full numerical solution (solid lines) for $\alpha_*=0$ and $\alpha_*=-0.04$ with a colour coding of $f=$ blue, $g=$ red, and $h=$ green.}
\end{figure}

\begin{figure}[H]
\begin{center}
\subfigure[The deviation for large-$r$ solutions for $\alpha_*=0$. ]{
  \includegraphics[scale=0.4]{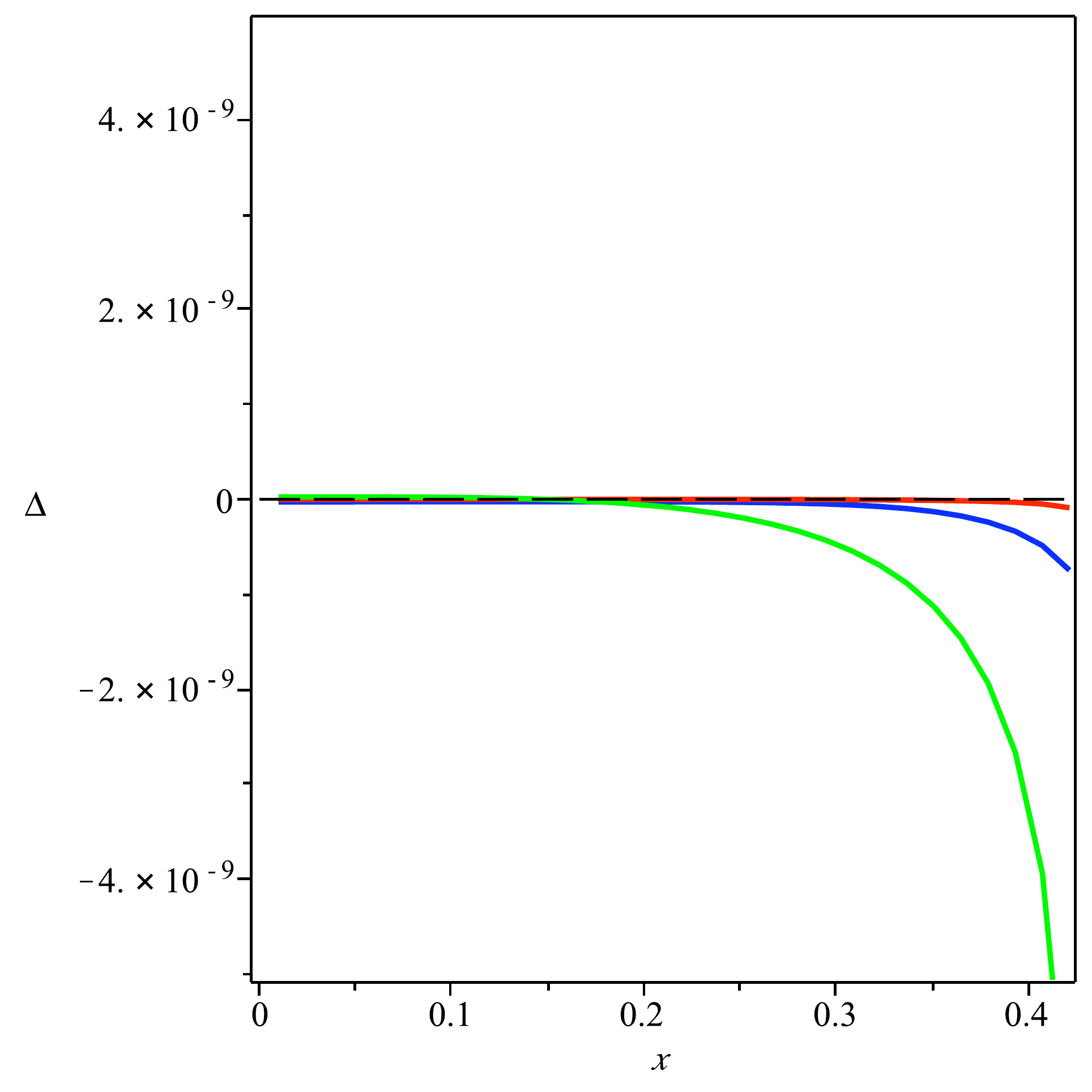}
 \label{subfig:dev_larger_alpha0}
}
\subfigure[The deviation of the large-$r$ solutions for $\alpha_*=-0.04$. ]{
 \includegraphics[scale=0.4]{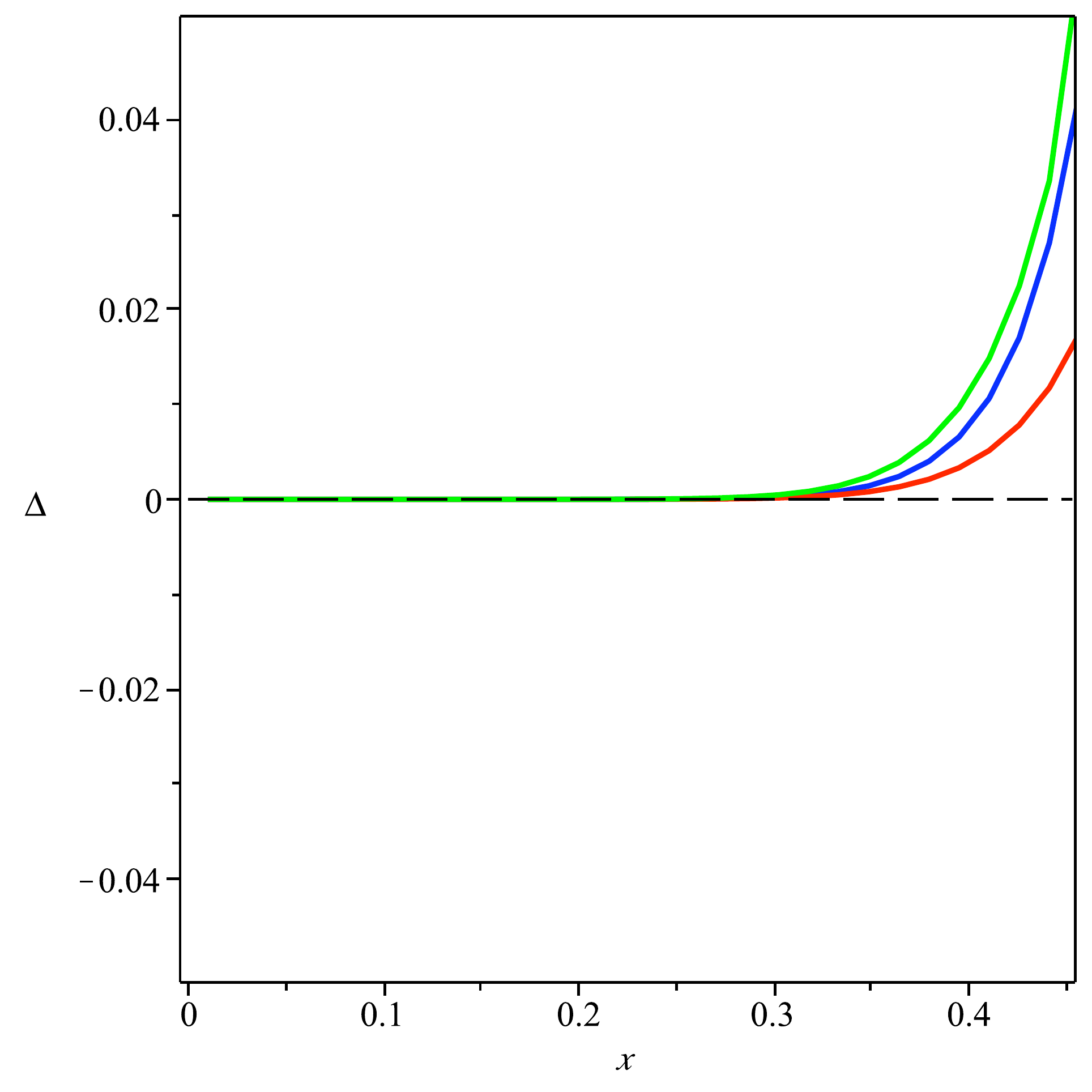}
 \label{subfig:dev_larger_alphan004}
}
\subfigure[The deviation of the near-$r_0$ solutions for $\alpha_*=0$. ]{
  \includegraphics[scale=0.4]{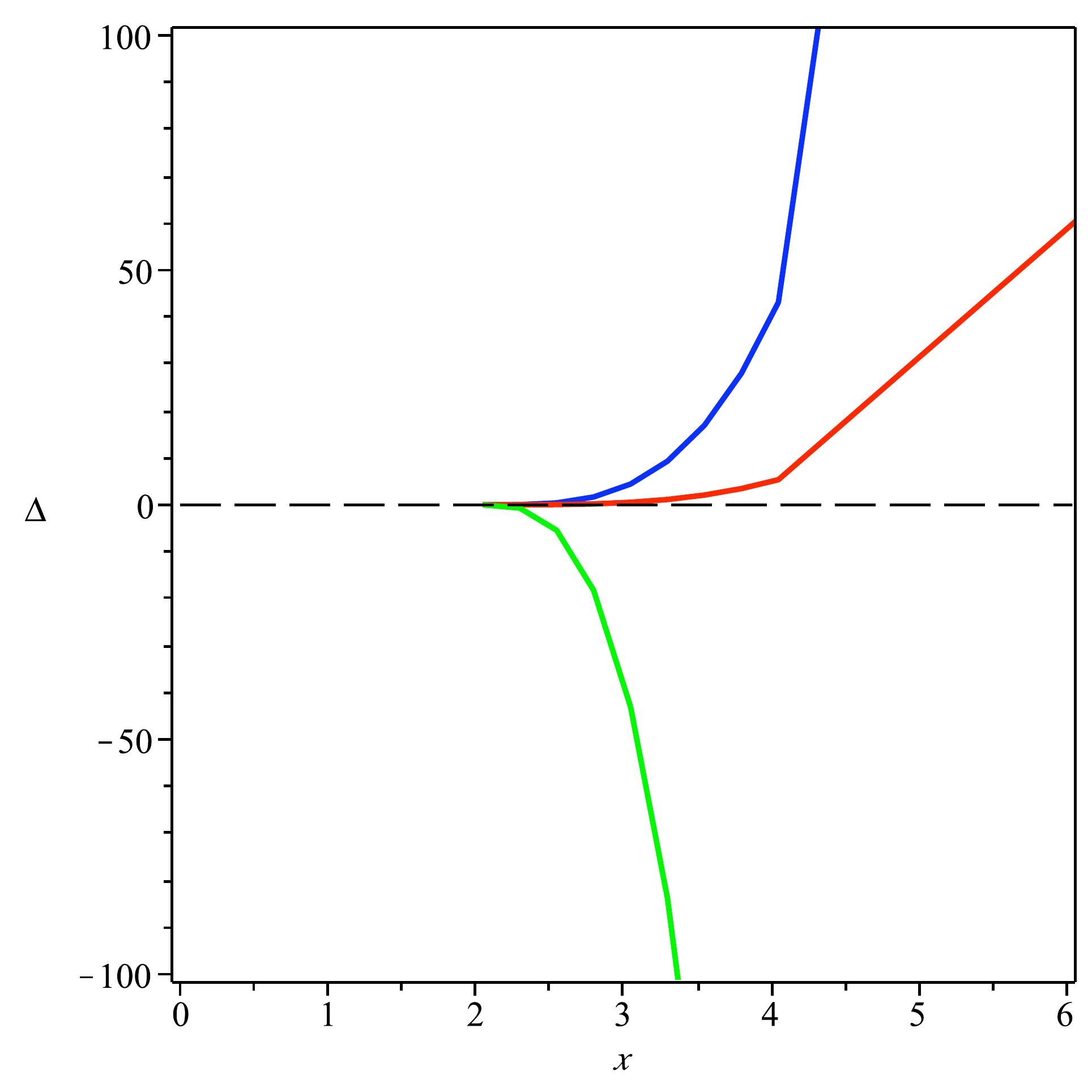}
 \label{subfig:dev_nearr0_alpha0}
}
\subfigure[The deviation of the near-$r_0$ solutions for $\alpha_*=-0.04$. ]{
 \includegraphics[scale=0.4]{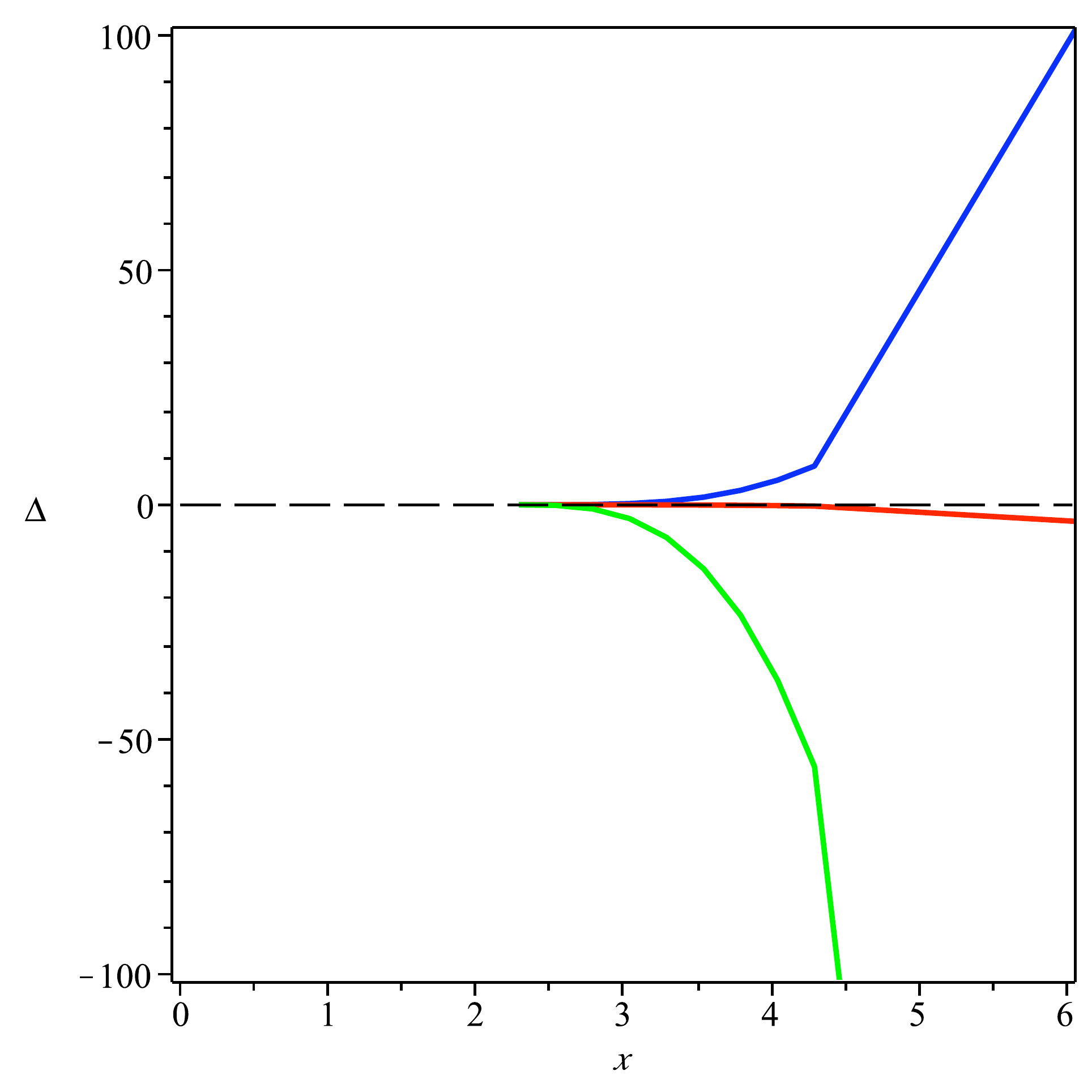}
 \label{subfig:dev_nearr0_alphan004}
}
\end{center}
\caption{ The deviation, $\Delta = F_{\text{ps}}(x)-F_{\text{num}}(x)$, of the large-$r$ and near-$r_0$ power-series solution with the full numerical solution for $\alpha_*=0$ and $\alpha_*=-0.04$ with a colour coding of $f=$ blue, $g=$ red, and $h=$ green. The large-$r$ power-series is expanded up to $1/r^{10}$, and the near-$r_0$ power-series is expanded up to $(r-r_0)^3$.}
\end{figure}

\section{  Large-$r$ power-series coefficients}
\label{app:large_r_pow_series_coeff}

We present here the near-$r_0$ power series coefficients up to $(r-r_0)^2$ from Eq. \eqref{eqn:EGB_EH_fgh_pow_near_r0_ansatz}. 

{\allowdisplaybreaks
\begin{equation*}
A_1 = 
\sqrt{\frac{4 \ell^2 p^2}{r_0^4 C_0}}
\end{equation*}
}
{\allowdisplaybreaks
\begin{eqnarray*}
A_2 &=& \frac{A^{\text{N}}_2}{A^{\text{D}}_2} \\
\quad\\
A^{\text{N}}_2 &=&
5\,{{\it r_0}}^{10}{{\it \ell}}^{4}pC_{{0}}+ \left( 2\,{{\it \ell}}^{5}
\sqrt {C_{{0}}}-68\,{{\it \ell}}^{3}{p}^{2}{C_{{0}}}^{3/2}\alpha
 \right) {{\it r_0}}^{9} \\ &&
 + \left( -96\,{{\it \ell}}^{4}pC_{{0}}\alpha-{{
\it \ell}}^{6}p+336\,{{\it \ell}}^{2}{p}^{3}{C_{{0}}}^{2}{\alpha}^{2}
 \right) {{\it r_0}}^{8} \\ &&
 + \left( -704\,{\it \ell}\,{p}^{4}{C_{{0}}}^{5/2
}{\alpha}^{3}-20\,{{\it \ell}}^{5}{p}^{2}\alpha\,\sqrt {C_{{0}}}+672\,{
{\it \ell}}^{3}{p}^{2}{C_{{0}}}^{3/2}{\alpha}^{2}-16\,{{\it \ell}}^{5}
\alpha\,\sqrt {C_{{0}}} \right) {{\it r_0}}^{7} \\ &&
+ \left( -24\,{{\it \ell}
}^{6}p\alpha+512\,{\alpha}^{4}{p}^{5}{C_{{0}}}^{3}-1280\,{{\it \ell}}^{
2}{p}^{3}{C_{{0}}}^{2}{\alpha}^{3}+224\,{{\it \ell}}^{4}{p}^{3}C_{{0}}{
\alpha}^{2}-768\,{{\it \ell}}^{4}pC_{{0}}{\alpha}^{2} \right) {{\it r_0}
}^{6} \\ &&
+ \left( -768\,{\alpha}^{2}{{\it \ell}}^{5}\sqrt {C_{{0}}}+8448\,{
{\it \ell}}^{3}{p}^{2}{C_{{0}}}^{3/2}{\alpha}^{3}+736\,{{\it \ell}}^{5}{
p}^{2}{\alpha}^{2}\sqrt {C_{{0}}}-768\,{\alpha}^{3}{{\it \ell}}^{3}{p}^
{4}{C_{{0}}}^{3/2} \right) {{\it r_0}}^{5} \\ &&
+ \left( 1024\,{\alpha}^{4}{{
\it \ell}}^{2}{p}^{5}{C_{{0}}}^{2}+12288\,{{\it \ell}}^{4}pC_{{0}}{
\alpha}^{3}-20480\,{\alpha}^{4}{{\it \ell}}^{2}{p}^{3}{C_{{0}}}^{2}-
5760\,{\alpha}^{3}{{\it \ell}}^{4}{p}^{3}C_{{0}}+16\,{\alpha}^{2}{{\it 
\ell}}^{6}{p}^{3} \right) {{\it r_0}}^{4} \\ &&
+ \left( -2816\,{{\it \ell}}^{5}
{p}^{2}{\alpha}^{3}\sqrt {C_{{0}}}+14336\,{\alpha}^{4}{{\it \ell}}^{3}{
p}^{4}{C_{{0}}}^{3/2}-36864\,{\alpha}^{4}{{\it \ell}}^{3}{p}^{2}{C_{{0}
}}^{3/2}-64\,{\alpha}^{3}{{\it \ell}}^{5}{p}^{4}\sqrt {C_{{0}}}
 \right) {{\it r_0}}^{3} \\ &&
 + \left( 14336\,{\alpha}^{4}{{\it \ell}}^{4}{p}^
{3}C_{{0}}-1536\,{\alpha}^{4}{{\it \ell}}^{4}{p}^{5}C_{{0}}-1408\,{
\alpha}^{3}{{\it \ell}}^{6}{p}^{3}+2048\,{\alpha}^{3}{{\it \ell}}^{6}p
 \right) {{\it r_0}}^{2} \\ &&
 + \left( 2048\,{\alpha}^{4}{{\it \ell}}^{5}{p}^{
4}\sqrt {C_{{0}}}-20480\,{\alpha}^{4}{{\it \ell}}^{5}{p}^{2}\sqrt {C_{{0
}}} \right) {\it r_0}+10240\,{\alpha}^{4}{{\it \ell}}^{6}{p}^{3} \\
\\
A^{\text{D}}_2 &=&
{{\it r_0}}^{5}{C_{{0}}}^{3/2} \left( {{\it r_0}}^{2}+16\,\alpha
 \right)  \left( 4\,{\it r_0}\,\alpha\,p\sqrt {C_{{0}}}+8\,\alpha\,{
\it \ell}-{\it \ell}\,{{\it r_0}}^{2} \right)  ( 64\,{\alpha}^{2}{
\it \ell}\,p\sqrt {C_{{0}}}{\it r_0} \\ &&
+16\,{p}^{2}C_{{0}}{\alpha}^{2}{{
\it r_0}}^{2}-16\,{\alpha}^{2}{{\it \ell}}^{2}{p}^{2}-16\,{{\it r_0}}^{2}
{{\it \ell}}^{2}\alpha-8\,{\it \ell}\,{{\it r_0}}^{3}\alpha\,p\sqrt {C_{{0
}}}+{{\it r_0}}^{4}{{\it \ell}}^{2} ) 
\end{eqnarray*}
}

{\allowdisplaybreaks
\begin{eqnarray*}
B_1 &=& \frac{B^{\text{N}}_1}{B^{\text{D}}_1} \\
\\
B^{\text{N}}_1 &=&
-4\,B_{{0}}{{\it \ell}}^{7}\sqrt {C_{{0}}}{{\it r_0}}^{13}+64\,B_{{0}}
\alpha\,{{\it \ell}}^{6}pC_{{0}}{{\it r_0}}^{12}-24\,B_{{0}}{{\it \ell}}^
{5}\alpha\,\sqrt {C_{{0}}} \left( -4\,{{\it \ell}}^{2}+{{\it \ell}}^{2}{
p}^{2}+8\,{p}^{2}C_{{0}}\alpha \right) {{\it r_0}}^{11} \\ &&
-32\,B_{{0}}{
\alpha}^{2}{{\it \ell}}^{4}pC_{{0}} \left( -15\,{{\it \ell}}^{2}{p}^{2}+
12\,{{\it \ell}}^{2}+80\,{p}^{2}C_{{0}}\alpha \right) {{\it r_0}}^{10} \\ &&
+
64\,B_{{0}}{\alpha}^{2}{{\it \ell}}^{3}\sqrt {C_{{0}}} \left( 17\,{{
\it \ell}}^{4}{p}^{2}+400\,{p}^{4}{C_{{0}}}^{2}{\alpha}^{2}+8\,{{\it 
\ell}}^{4}-240\,{{\it \ell}}^{2}{p}^{2}C_{{0}}\alpha-60\,\alpha\,{{\it 
\ell}}^{2}{p}^{4}C_{{0}} \right) {{\it r_0}}^{9} \\ && 
-1024\,B_{{0}}{\alpha}^{
3}{{\it \ell}}^{2}pC_{{0}} \left( -15\,\alpha\,{{\it \ell}}^{2}{p}^{4}C_
{{0}}+34\,{{\it \ell}}^{4}+96\,{p}^{4}{C_{{0}}}^{2}{\alpha}^{2}-180\,{{
\it \ell}}^{2}{p}^{2}C_{{0}}\alpha+16\,{{\it \ell}}^{4}{p}^{2} \right) {
{\it r_0}}^{8} \\ &&
+256\,B_{{0}}{\alpha}^{3}\sqrt {C_{{0}}}{\it \ell}\,
 ( 704\,{\alpha}^{3}{p}^{6}{C_{{0}}}^{3}+1856\,{p}^{2}\alpha\,{{
\it \ell}}^{4}C_{{0}}-96\,{{\it \ell}}^{6}-52\,{{\it \ell}}^{6}{p}^{2}-
120\,{\alpha}^{2}{{\it \ell}}^{2}{p}^{6}{C_{{0}}}^{2}  \\ &&
-3360\,{\alpha}^{2
}{{\it \ell}}^{2}{p}^{4}{C_{{0}}}^{2}+344\,\alpha\,{{\it \ell}}^{4}{p}^{
4}C_{{0}}+3\,{{\it \ell}}^{6}{p}^{4} ) {{\it r_0}}^{7} \\ &&
-1024\,B_{{0
}}{\alpha}^{4}pC_{{0}} ( 176\,\alpha\,{{\it \ell}}^{4}{p}^{4}C_{{0
}}-24\,{\alpha}^{2}{{\it \ell}}^{2}{p}^{6}{C_{{0}}}^{2}+9\,{{\it \ell}}^
{6}{p}^{4}+128\,{\alpha}^{3}{p}^{6}{C_{{0}}}^{3}-480\,{{\it \ell}}^{6} \\ &&
-
1824\,{\alpha}^{2}{{\it \ell}}^{2}{p}^{4}{C_{{0}}}^{2}-124\,{{\it \ell}}
^{6}{p}^{2}+2624\,{p}^{2}\alpha\,{{\it \ell}}^{4}C_{{0}} ) {{\it 
r_0}}^{6} \\ && 
-1024\,B_{{0}}{\alpha}^{4}{\it \ell}\,\sqrt {C_{{0}}} ( -
128\,{{\it \ell}}^{6}+240\,\alpha\,{{\it \ell}}^{4}{p}^{4}C_{{0}}+1536\,
{\alpha}^{3}{p}^{6}{C_{{0}}}^{3}+3456\,{p}^{2}\alpha\,{{\it \ell}}^{4}C
_{{0}}-16\,{{\it \ell}}^{6}{p}^{2} \\ &&
-6784\,{\alpha}^{2}{{\it \ell}}^{2}{p}
^{4}{C_{{0}}}^{2}-36\,\alpha\,{{\it \ell}}^{4}{p}^{6}C_{{0}}-16\,{
\alpha}^{2}{{\it \ell}}^{2}{p}^{6}{C_{{0}}}^{2}+19\,{{\it \ell}}^{6}{p}^
{4} ) {{\it r_0}}^{5} \\ &&
+16384\,B_{{0}}{\alpha}^{5}{{\it \ell}}^{2}pC
_{{0}} ( 12\,{{\it \ell}}^{4}{p}^{2}+16\,{C_{{0}}}^{2}{p}^{6}{
\alpha}^{2}-96\,{{\it \ell}}^{4}-416\,{p}^{4}{C_{{0}}}^{2}{\alpha}^{2}+
672\,{{\it \ell}}^{2}{p}^{2}C_{{0}}\alpha \\ &&
-3\,C_{{0}}{p}^{6}{{\it \ell}}^
{2}\alpha+9\,{p}^{4}{{\it \ell}}^{4}-44\,\alpha\,{{\it \ell}}^{2}{p}^{4}
C_{{0}} ) {{\it r_0}}^{4}+2048\,B_{{0}}{\alpha}^{5}{{\it \ell}}^{3
}{p}^{2}\sqrt {C_{{0}}} ( -1152\,{{\it \ell}}^{2}{p}^{2}C_{{0}}
\alpha \\ &&
+44\,{{\it \ell}}^{4}{p}^{2}-6144\,{p}^{2}{C_{{0}}}^{2}{\alpha}^{
2}-3\,{p}^{4}{{\it \ell}}^{4}+3072\,{{\it \ell}}^{2}C_{{0}}\alpha+1024\,
{p}^{4}{C_{{0}}}^{2}{\alpha}^{2}+128\,{{\it \ell}}^{4} \\ && 
-120\,\alpha\,{{
\it \ell}}^{2}{p}^{4}C_{{0}} ) {{\it r_0}}^{3} \\ &&
-8192\,B_{{0}}{
\alpha}^{6}{{\it \ell}}^{4}{p}^{3}C_{{0}} \left( -3\,{{\it \ell}}^{2}{p}
^{4}+256\,{{\it \ell}}^{2}+28\,{{\it \ell}}^{2}{p}^{2}+16\,\alpha\,{p}^{
4}C_{{0}}-640\,{p}^{2}C_{{0}}\alpha+1024\,\alpha\,C_{{0}} \right) {{
\it r_0}}^{2} \\ &&
-16384\,B_{{0}}{\alpha}^{6}{{\it \ell}}^{5}{p}^{4}\sqrt {C_
{{0}}} \left( -256\,\alpha\,C_{{0}}+32\,{p}^{2}C_{{0}}\alpha-3\,{{\it 
\ell}}^{2}{p}^{2}-8\,{{\it \ell}}^{2} \right) {\it r_0}-524288\,B_{{0}}{
\alpha}^{7}{p}^{5}C_{{0}}{{\it \ell}}^{6} \\
\\
B^{\text{D}}_1 &=&
{\it r_0}\,p \left( {{\it r_0}}^{2}+16\,\alpha \right)  \left( 4\,{\it 
r_0}\,\alpha\,p\sqrt {C_{{0}}}+8\,\alpha\,{\it \ell}-{\it \ell}\,{{\it r_0
}}^{2} \right) ^{2} ( 64\,{\alpha}^{2}{\it \ell}\,p\sqrt {C_{{0}}}
{\it r_0}+16\,{p}^{2}C_{{0}}{\alpha}^{2}{{\it r_0}}^{2} \\ &&
-16\,{\alpha}^{2}
{{\it \ell}}^{2}{p}^{2}-16\,{{\it r_0}}^{2}{{\it \ell}}^{2}\alpha-8\,{
\it \ell}\,{{\it r_0}}^{3}\alpha\,p\sqrt {C_{{0}}}+{{\it r_0}}^{4}{{\it 
\ell}}^{2} ) ^{2}C_{{0}}
\end{eqnarray*}
}

{\allowdisplaybreaks
\begin{eqnarray*}
B_2 &=& \frac{B^{\text{N}}_2}{B^{\text{D}}_2} \\
\\
B^{\text{N}}_2 &=&
4\,{{\it \ell}}^{5}pC_{{0}}B_{{0}}{{\it r_0}}^{12}+{{\it \ell}}^{4}\sqrt 
{C_{{0}}} \left( {{\it \ell}}^{2}{p}^{2}+4\,{{\it \ell}}^{2}+16\,{p}^{2}
C_{{0}}\alpha \right) B_{{0}}{{\it r_0}}^{11} \\ &&
-2\,{{\it \ell}}^{3}p
 \left( 448\,{p}^{2}{C_{{0}}}^{2}{\alpha}^{2}+{{\it \ell}}^{4}-34\,{{
\it \ell}}^{2}{p}^{2}C_{{0}}\alpha-24\,{{\it \ell}}^{2}C_{{0}}\alpha
 \right) B_{{0}}{{\it r_0}}^{10} \\ &&
 +8\,\alpha\,{{\it \ell}}^{2}\sqrt {C_{{0
}}} \left( 15\,{{\it \ell}}^{4}{p}^{2}-248\,{{\it \ell}}^{2}{p}^{2}C_{{0
}}\alpha-114\,\alpha\,{{\it \ell}}^{2}{p}^{4}C_{{0}}+4\,{{\it \ell}}^{4}
+832\,{p}^{4}{C_{{0}}}^{2}{\alpha}^{2} \right) B_{{0}}{{\it r_0}}^{9} \\ &&
-4
\,\alpha\,{\it \ell}\,p ( 20\,{{\it \ell}}^{6}-944\,{\alpha}^{2}{{
\it \ell}}^{2}{p}^{4}{C_{{0}}}^{2}+3\,{{\it \ell}}^{6}{p}^{2}-3392\,{p}^
{2}{{\it \ell}}^{2}{C_{{0}}}^{2}{\alpha}^{2} +640\,{{\it \ell}}^{4}C_{{0}
}\alpha \\ &&
+560\,{p}^{2}\alpha\,{{\it \ell}}^{4}C_{{0}}+4864\,{p}^{4}{
\alpha}^{3}{C_{{0}}}^{3} ) B_{{0}}{{\it r_0}}^{8} \\ &&
+32\,{\alpha}^{2
}\sqrt {C_{{0}}} ( 240\,{p}^{2}\alpha\,{{\it \ell}}^{4}C_{{0}}-576
\,{\alpha}^{2}{{\it \ell}}^{2}{p}^{4}{C_{{0}}}^{2}-160\,{\alpha}^{2}{{
\it \ell}}^{2}{p}^{6}{C_{{0}}}^{2}+340\,\alpha\,{{\it \ell}}^{4}{p}^{4}C
_{{0}}-64\,{{\it \ell}}^{6} \\ &&
+640\,{\alpha}^{3}{p}^{6}{C_{{0}}}^{3}+8\,{{
\it \ell}}^{6}{p}^{2}+{{\it \ell}}^{6}{p}^{4} ) B_{{0}}{{\it r_0}}^
{7} \\ &&
-256\,{\alpha}^{2}{\it \ell}\,p ( 320\,{p}^{4}{\alpha}^{3}{C_{{0
}}}^{3}+{{\it \ell}}^{6}{p}^{2}-592\,{p}^{2}{{\it \ell}}^{2}{C_{{0}}}^{2
}{\alpha}^{2}+3\,{{\it \ell}}^{6}+30\,{\alpha}^{2}{{\it \ell}}^{2}{p}^{4
}{C_{{0}}}^{2} \\ &&
+74\,{p}^{2}\alpha\,{{\it \ell}}^{4}C_{{0}}+48\,{{\it \ell
}}^{4}C_{{0}}\alpha-2\,\alpha\,{{\it \ell}}^{4}{p}^{4}C_{{0}} ) B
_{{0}}{{\it r_0}}^{6}+128\,{\alpha}^{3}\sqrt {C_{{0}}} ( -140\,{{
\it \ell}}^{6}{p}^{2}+43\,{{\it \ell}}^{6}{p}^{4} \\ &&
-192\,{{\it \ell}}^{6}-
14\,\alpha\,{{\it \ell}}^{4}{p}^{6}C_{{0}}+3456\,{p}^{2}\alpha\,{{\it 
\ell}}^{4}C_{{0}}+1416\,\alpha\,{{\it \ell}}^{4}{p}^{4}C_{{0}}-224\,{
\alpha}^{2}{{\it \ell}}^{2}{p}^{6}{C_{{0}}}^{2} \\ &&
+1536\,{\alpha}^{3}{p}^{
6}{C_{{0}}}^{3}-7552\,{\alpha}^{2}{{\it \ell}}^{2}{p}^{4}{C_{{0}}}^{2}
 ) B_{{0}}{{\it r_0}}^{5} \\ &&
 +64\,{\alpha}^{3}{\it \ell}\,p ( -
864\,\alpha\,{{\it \ell}}^{4}{p}^{4}C_{{0}}-37888\,{p}^{2}{{\it \ell}}^{
2}{C_{{0}}}^{2}{\alpha}^{2}-36\,{{\it \ell}}^{6}{p}^{2}+4352\,{p}^{2}
\alpha\,{{\it \ell}}^{4}C_{{0}} \\ &&
+6144\,{{\it \ell}}^{4}C_{{0}}\alpha-5760
\,{\alpha}^{2}{{\it \ell}}^{2}{p}^{4}{C_{{0}}}^{2}+64\,{{\it \ell}}^{6}+
3\,{{\it \ell}}^{6}{p}^{4}+24576\,{p}^{4}{\alpha}^{3}{C_{{0}}}^{3}
 ) B_{{0}}{{\it r_0}}^{4} \\ &&
 -256\,{\alpha}^{4}{{\it \ell}}^{2}{p}^{2}
\sqrt {C_{{0}}} ( -15360\,{p}^{2}{C_{{0}}}^{2}{\alpha}^{2}+236\,{
{\it \ell}}^{4}{p}^{2}+2560\,{{\it \ell}}^{2}{p}^{2}C_{{0}}\alpha+7680\,
{{\it \ell}}^{2}C_{{0}}\alpha \\ &&
+21\,{p}^{4}{{\it \ell}}^{4}+768\,{p}^{4}{C
_{{0}}}^{2}{\alpha}^{2}-128\,{{\it \ell}}^{4}-560\,\alpha\,{{\it \ell}}^
{2}{p}^{4}C_{{0}} ) B_{{0}}{{\it r_0}}^{3} \\ &&
-512\,{\alpha}^{4}{{
\it \ell}}^{3}p ( 768\,{{\it \ell}}^{2}{p}^{2}C_{{0}}\alpha-128\,{{
\it \ell}}^{4}+1536\,{p}^{4}{C_{{0}}}^{2}{\alpha}^{2}-512\,\alpha\,{{
\it \ell}}^{2}{p}^{4}C_{{0}}+33\,{p}^{4}{{\it \ell}}^{4} \\ &&
-6144\,{p}^{2}{C
_{{0}}}^{2}{\alpha}^{2} ) B_{{0}}{{\it r_0}}^{2} \\ &&
-4096\,{\alpha}^{
5}{{\it \ell}}^{4}{p}^{2}\sqrt {C_{{0}}} \left( 160\,{{\it \ell}}^{2}-28
\,{{\it \ell}}^{2}{p}^{2}+192\,{p}^{2}C_{{0}}\alpha-15\,{{\it \ell}}^{2}
{p}^{4} \right) B_{{0}}{\it r_0} \\ &&
+40960\,{\alpha}^{5}{{\it \ell}}^{7}{p}^
{3} \left( 3\,{p}^{2}+8 \right) B_{{0}} \\
\\
B^{\text{D}}_2 &=&
{{\it r_0}}^{3}{C_{{0}}}^{3/2}{p}^{2} ( 64\,{\alpha}^{2}{\it \ell}
\,p\sqrt {C_{{0}}}{\it r_0}+16\,{p}^{2}C_{{0}}{\alpha}^{2}{{\it r_0}}^{2
}-16\,{\alpha}^{2}{{\it \ell}}^{2}{p}^{2}-16\,{{\it r_0}}^{2}{{\it \ell}}
^{2}\alpha \\ &&
- 8\,{\it \ell}\,{{\it r_0}}^{3}\alpha\,p\sqrt {C_{{0}}}+{{\it 
r_0}}^{4}{{\it \ell}}^{2} )  \left( 4\,{\it r_0}\,\alpha\,p\sqrt {C
_{{0}}}+8\,\alpha\,{\it \ell}-{\it \ell}\,{{\it r_0}}^{2} \right) ^{2}
 \left( {{\it r_0}}^{2}+16\,\alpha \right) ^{2} 
\end{eqnarray*}
}

{\allowdisplaybreaks
\begin{eqnarray*}
C_1 &=& \frac{C^{\text{N}}_1}{C^{\text{D}}_1} \\
\\
C^{\text{N}}_1 &=&
2\,{{\it \ell}}^{3}p \left( {{\it \ell}}^{2}+16\,\alpha\,C_{{0}}
 \right) {{\it r_0}}^{8}-16\,\alpha\,{{\it \ell}}^{2}\sqrt {C_{{0}}}
 \left( -2\,{{\it \ell}}^{2}+24\,{p}^{2}C_{{0}}\alpha-{{\it \ell}}^{2}{p
}^{2} \right) {{\it r_0}}^{7} \\ &&
+16\,\alpha\,{\it \ell}\,p \left( -16\,{{
\it \ell}}^{2}C_{{0}}\alpha-14\,{{\it \ell}}^{2}{p}^{2}C_{{0}}\alpha+3\,
{{\it \ell}}^{4}+96\,{p}^{2}{C_{{0}}}^{2}{\alpha}^{2} \right) {{\it r_0}
}^{6} \\ &&
-256\,{\alpha}^{2}\sqrt {C_{{0}}} \left( -2\,{{\it \ell}}^{4}+8\,{
p}^{4}{C_{{0}}}^{2}{\alpha}^{2}-2\,\alpha\,{{\it \ell}}^{2}{p}^{4}C_{{0
}}+3\,{{\it \ell}}^{4}{p}^{2}-2\,{{\it \ell}}^{2}{p}^{2}C_{{0}}\alpha
 \right) {{\it r_0}}^{5} \\ &&
 +32\,{\alpha}^{2}{{\it \ell}}^{3}p \left( -{{
\it \ell}}^{2}{p}^{2}+104\,{p}^{2}C_{{0}}\alpha-128\,\alpha\,C_{{0}}
 \right) {{\it r_0}}^{4} \\ &&
 -512\,{\alpha}^{3}{{\it \ell}}^{2}{p}^{2}\sqrt {
C_{{0}}} \left( -16\,\alpha\,C_{{0}}+8\,{p}^{2}C_{{0}}\alpha+3\,{{\it 
\ell}}^{2}-{{\it \ell}}^{2}{p}^{2} \right) {{\it r_0}}^{3} \\ &&
+256\,{\alpha}^
{3}{{\it \ell}}^{3}p \left( 11\,{{\it \ell}}^{2}{p}^{2}-16\,{{\it \ell}}^
{2}+48\,{p}^{2}C_{{0}}\alpha \right) {{\it r_0}}^{2} \\ &&
-10240\,{\alpha}^{4
}{{\it \ell}}^{4}{p}^{2}\sqrt {C_{{0}}} \left( -2+p \right)  \left( p+2
 \right) {\it r_0}-20480\,{\alpha}^{4}{{\it \ell}}^{5}{p}^{3} \\
\\
C^{\text{D}}_1 &=&
{{\it r_0}}^{3}p ( 64\,{\alpha}^{2}{\it \ell}\,p\sqrt {C_{{0}}}{
\it r_0}+16\,{p}^{2}C_{{0}}{\alpha}^{2}{{\it r_0}}^{2}-16\,{\alpha}^{2}{
{\it \ell}}^{2}{p}^{2}-16\,{{\it r_0}}^{2}{{\it \ell}}^{2}\alpha \\ &&
-8\,{\it 
\ell}\,{{\it r_0}}^{3}\alpha\,p\sqrt {C_{{0}}}+{{\it r_0}}^{4}{{\it \ell}}
^{2} )  \left( 4\,{\it r_0}\,\alpha\,p\sqrt {C_{{0}}}+8\,\alpha\,
{\it \ell}-{\it \ell}\,{{\it r_0}}^{2} \right)  \left( {{\it r_0}}^{2}+16
\,\alpha \right) 
\end{eqnarray*}
}

{\allowdisplaybreaks
\begin{eqnarray*}
C_2 &=& \frac{C^{\text{N}}_2}{C^{\text{D}}_2} \\
\\
C^{\text{N}}_2 &=&
-2\,{{\it \ell}}^{9}{{\it r_0}}^{22}pC_{{0}}+2\,{{\it \ell}}^{8}\sqrt {C_
{{0}}} \left( 144\,{p}^{2}C_{{0}}\alpha+2\,{{\it \ell}}^{2}+5\,{{\it 
\ell}}^{2}{p}^{2} \right) {{\it r_0}}^{21} \\ &&
-2\,{{\it \ell}}^{7}p \left( 8
\,{{\it \ell}}^{2}C_{{0}}\alpha+{{\it \ell}}^{4}+4032\,{p}^{2}{C_{{0}}}^
{2}{\alpha}^{2}-54\,{{\it \ell}}^{2}{p}^{2}C_{{0}}\alpha \right) {{\it 
r_0}}^{20} \\ &&
+8\,\alpha\,{{\it \ell}}^{6}\sqrt {C_{{0}}} \left( -942\,
\alpha\,{{\it \ell}}^{2}{p}^{4}C_{{0}}+13440\,{p}^{4}{C_{{0}}}^{2}{
\alpha}^{2}-328\,{{\it \ell}}^{2}{p}^{2}C_{{0}}\alpha-20\,{{\it \ell}}^{
4}+71\,{{\it \ell}}^{4}{p}^{2} \right) {{\it r_0}}^{19} \\ &&
-4\,\alpha\,{{
\it \ell}}^{5}p ( -39\,{{\it \ell}}^{6}{p}^{2}+209664\,{p}^{4}{
\alpha}^{3}{C_{{0}}}^{3}+6456\,{p}^{2}\alpha\,{{\it \ell}}^{4}C_{{0}}-
14272\,{p}^{2}{{\it \ell}}^{2}{C_{{0}}}^{2}{\alpha}^{2} \\ &&
-2464\,{{\it \ell
}}^{4}C_{{0}}\alpha-4\,{{\it \ell}}^{6}-27216\,{\alpha}^{2}{{\it \ell}}^
{2}{p}^{4}{C_{{0}}}^{2} ) {{\it r_0}}^{18} \\ &&
+16\,{\alpha}^{2}{{\it 
\ell}}^{4}\sqrt {C_{{0}}} ( 258048\,{\alpha}^{3}{p}^{6}{C_{{0}}}^{
3}-96\,{{\it \ell}}^{6}-292\,{{\it \ell}}^{6}{p}^{2}-34496\,{\alpha}^{2}
{{\it \ell}}^{2}{p}^{4}{C_{{0}}}^{2}-223\,{{\it \ell}}^{6}{p}^{4} \\ &&
-49392
\,{\alpha}^{2}{{\it \ell}}^{2}{p}^{6}{C_{{0}}}^{2}+27720\,\alpha\,{{
\it \ell}}^{4}{p}^{4}C_{{0}}-14112\,{p}^{2}\alpha\,{{\it \ell}}^{4}C_{{0
}} ) {{\it r_0}}^{17} \\ &&
-64\,{\alpha}^{2}{{\it \ell}}^{3}p ( -
716\,{p}^{4}{{\it \ell}}^{6}\alpha\,C_{{0}}+61704\,{p}^{4}{{\it \ell}}^{
4}{\alpha}^{2}{C_{{0}}}^{2}-2280\,{p}^{2}{{\it \ell}}^{6}\alpha\,C_{{0}
}-52080\,{p}^{6}{{\it \ell}}^{2}{\alpha}^{3}{C_{{0}}}^{3} \\ &&
-11\,{p}^{2}{{
\it \ell}}^{8}-288\,{{\it \ell}}^{6}\alpha\,C_{{0}}+204288\,{p}^{6}{
\alpha}^{4}{C_{{0}}}^{4}-46144\,{p}^{4}{{\it \ell}}^{2}{\alpha}^{3}{C_{
{0}}}^{3}-32\,{{\it \ell}}^{8} \\ && 
-41504\,{p}^{2}{{\it \ell}}^{4}{\alpha}^{2
}{C_{{0}}}^{2} ) {{\it r_0}}^{16}+128\,{\alpha}^{3}{{\it \ell}}^{2
}\sqrt {C_{{0}}} ( -65184\,{\alpha}^{3}{{\it \ell}}^{2}{p}^{8}{C_{
{0}}}^{3}+704\,{{\it \ell}}^{8} \\ && 
+160176\,{\alpha}^{2}{{\it \ell}}^{4}{p}^
{6}{C_{{0}}}^{2}-12528\,{p}^{4}{{\it \ell}}^{6}\alpha\,C_{{0}}-1836\,{p
}^{2}{{\it \ell}}^{8}-4352\,{p}^{2}{{\it \ell}}^{6}\alpha\,C_{{0}}+439\,
{{\it \ell}}^{8}{p}^{4} \\ &&
-68992\,{p}^{6}{{\it \ell}}^{2}{\alpha}^{3}{C_{{0
}}}^{3}-3264\,\alpha\,{{\it \ell}}^{6}{p}^{6}C_{{0}}-140992\,{p}^{4}{{
\it \ell}}^{4}{\alpha}^{2}{C_{{0}}}^{2}+202752\,{\alpha}^{4}{p}^{8}{C_{
{0}}}^{4} ) {{\it r_0}}^{15} \\ &&
-64\,{\alpha}^{3}{\it \ell}\,p ( 
-1144320\,{p}^{4}{{\it \ell}}^{4}{\alpha}^{3}{C_{{0}}}^{3}-211968\,{p}^
{2}{{\it \ell}}^{6}{\alpha}^{2}{C_{{0}}}^{2}-185344\,{p}^{6}{{\it \ell}}
^{2}{\alpha}^{4}{C_{{0}}}^{4} \\ &&
-182016\,{p}^{8}{{\it \ell}}^{2}{\alpha}^{
4}{C_{{0}}}^{4}+1008768\,{p}^{6}{{\it \ell}}^{4}{\alpha}^{3}{C_{{0}}}^{
3}+51456\,{{\it \ell}}^{8}\alpha\,C_{{0}}-123200\,{p}^{4}{{\it \ell}}^{6
}{\alpha}^{2}{C_{{0}}}^{2} \\ &&
+460800\,{p}^{8}{\alpha}^{5}{C_{{0}}}^{5}-
39920\,{p}^{6}{{\it \ell}}^{6}{\alpha}^{2}{C_{{0}}}^{2}+660\,{p}^{2}{{
\it \ell}}^{10}-21\,{p}^{4}{{\it \ell}}^{10}+29320\,{p}^{4}{{\it \ell}}^{
8}\alpha\,C_{{0}} \\ &&
-93856\,{p}^{2}{{\it \ell}}^{8}\alpha\,C_{{0}}
 ) {{\it r_0}}^{14}+256\,{\alpha}^{4}\sqrt {C_{{0}}} ( -
275904\,{p}^{4}{{\it \ell}}^{8}\alpha\,C_{{0}}+479616\,{\alpha}^{3}{{
\it \ell}}^{4}{p}^{8}{C_{{0}}}^{3}+512\,{{\it \ell}}^{10} \\ &&
-37744\,{\alpha
}^{2}{{\it \ell}}^{6}{p}^{8}{C_{{0}}}^{2}+207360\,{p}^{2}{{\it \ell}}^{8
}\alpha\,C_{{0}}-7772\,{p}^{4}{{\it \ell}}^{10}+57344\,{\alpha}^{5}{p}^
{10}{C_{{0}}}^{5} \\ &&
-591360\,{p}^{4}{{\it \ell}}^{6}{\alpha}^{2}{C_{{0}}}^
{2}-28416\,{\alpha}^{4}{{\it \ell}}^{2}{p}^{10}{C_{{0}}}^{4}+5600\,{p}^
{2}{{\it \ell}}^{10}-70464\,{p}^{6}{{\it \ell}}^{6}{\alpha}^{2}{C_{{0}}}
^{2} \\ &&
+85328\,\alpha\,{{\it \ell}}^{8}{p}^{6}C_{{0}}-669184\,{p}^{6}{{
\it \ell}}^{4}{\alpha}^{3}{C_{{0}}}^{3}-59\,{{\it \ell}}^{10}{p}^{6}+
22528\,{p}^{8}{{\it \ell}}^{2}{\alpha}^{4}{C_{{0}}}^{4} ) {{\it 
r_0}}^{13} \\ &&
+1024\,{\alpha}^{4}{\it \ell}\,p ( -36864\,{p}^{8}{\alpha
}^{5}{C_{{0}}}^{5}+20448\,{\alpha}^{3}{{\it \ell}}^{4}{p}^{8}{C_{{0}}}^
{3}+49804\,{p}^{4}{{\it \ell}}^{8}\alpha\,C_{{0}} \\ &&
-134400\,{p}^{8}{{\it 
\ell}}^{2}{\alpha}^{4}{C_{{0}}}^{4}-134\,{p}^{4}{{\it \ell}}^{10}+33792
\,{p}^{6}{{\it \ell}}^{4}{\alpha}^{3}{C_{{0}}}^{3}+866816\,{p}^{4}{{
\it \ell}}^{4}{\alpha}^{3}{C_{{0}}}^{3} \\ &&
-33728\,{p}^{2}{{\it \ell}}^{8}
\alpha\,C_{{0}}+189952\,{p}^{6}{{\it \ell}}^{2}{\alpha}^{4}{C_{{0}}}^{4
}+169\,\alpha\,{{\it \ell}}^{8}{p}^{6}C_{{0}}+425536\,{p}^{4}{{\it \ell}
}^{6}{\alpha}^{2}{C_{{0}}}^{2}-512\,{{\it \ell}}^{10} \\ &&
+512\,{\alpha}^{5}
{p}^{10}{C_{{0}}}^{5}-208\,{p}^{2}{{\it \ell}}^{10}-124768\,{p}^{6}{{
\it \ell}}^{6}{\alpha}^{2}{C_{{0}}}^{2}-462848\,{p}^{2}{{\it \ell}}^{6}{
\alpha}^{2}{C_{{0}}}^{2}-3072\,{{\it \ell}}^{8}\alpha\,C_{{0}} ) 
{{\it r_0}}^{12} \\ &&
+2048\,{\alpha}^{5}\sqrt {C_{{0}}} ( 12544\,{p}^{2
}{{\it \ell}}^{10}-766\,\alpha\,{{\it \ell}}^{8}{p}^{8}C_{{0}}-12288\,{p
}^{8}{{\it \ell}}^{2}{\alpha}^{4}{C_{{0}}}^{4}+45568\,{\alpha}^{4}{{
\it \ell}}^{2}{p}^{10}{C_{{0}}}^{4} \\ &&
+16384\,{\alpha}^{5}{p}^{10}{C_{{0}}
}^{5}+110208\,{p}^{4}{{\it \ell}}^{8}\alpha\,C_{{0}}-217288\,\alpha\,{{
\it \ell}}^{8}{p}^{6}C_{{0}}+1219584\,{p}^{4}{{\it \ell}}^{6}{\alpha}^{2
}{C_{{0}}}^{2} \\ &&
+198832\,{\alpha}^{2}{{\it \ell}}^{6}{p}^{8}{C_{{0}}}^{2}
-6144\,{{\it \ell}}^{10}-11072\,{\alpha}^{3}{{\it \ell}}^{4}{p}^{10}{C_{
{0}}}^{3}-1406976\,{p}^{6}{{\it \ell}}^{4}{\alpha}^{3}{C_{{0}}}^{3}+
13556\,{p}^{4}{{\it \ell}}^{10} \\ &&
+12288\,{p}^{2}{{\it \ell}}^{8}\alpha\,C_
{{0}}-669632\,{p}^{6}{{\it \ell}}^{6}{\alpha}^{2}{C_{{0}}}^{2}-89600\,{
\alpha}^{3}{{\it \ell}}^{4}{p}^{8}{C_{{0}}}^{3}+1411\,{{\it \ell}}^{10}{
p}^{6} ) {{\it r_0}}^{11} \\ &&
-1024\,{\alpha}^{5}{\it \ell}\,p ( 
358304\,{p}^{4}{{\it \ell}}^{8}\alpha\,C_{{0}}+28616\,\alpha\,{{\it \ell
}}^{8}{p}^{6}C_{{0}}+564736\,{p}^{2}{{\it \ell}}^{8}\alpha\,C_{{0}}+
98304\,{p}^{8}{\alpha}^{5}{C_{{0}}}^{5} \\ &&
-4816896\,{p}^{6}{{\it \ell}}^{2
}{\alpha}^{4}{C_{{0}}}^{4}-233472\,{{\it \ell}}^{8}\alpha\,C_{{0}}+
7557120\,{p}^{4}{{\it \ell}}^{4}{\alpha}^{3}{C_{{0}}}^{3}-589824\,{p}^{
8}{{\it \ell}}^{2}{\alpha}^{4}{C_{{0}}}^{4} \\ &&
-8192\,{\alpha}^{4}{{\it \ell
}}^{2}{p}^{10}{C_{{0}}}^{4}+159\,{{\it \ell}}^{10}{p}^{6}-1851008\,{p}^
{6}{{\it \ell}}^{6}{\alpha}^{2}{C_{{0}}}^{2}-3840\,{p}^{2}{{\it \ell}}^{
10}-5596\,{p}^{4}{{\it \ell}}^{10} \\ &&
+65536\,{p}^{2}{{\it \ell}}^{6}{\alpha
}^{2}{C_{{0}}}^{2}+40960\,{\alpha}^{5}{p}^{10}{C_{{0}}}^{5}+668800\,{
\alpha}^{3}{{\it \ell}}^{4}{p}^{8}{C_{{0}}}^{3}-219136\,{p}^{4}{{\it 
\ell}}^{6}{\alpha}^{2}{C_{{0}}}^{2} \\ &&
-6560\,{\alpha}^{2}{{\it \ell}}^{6}{p
}^{8}{C_{{0}}}^{2}+1024\,{{\it \ell}}^{10} -1544704\,{p}^{6}{{\it \ell}}^
{4}{\alpha}^{3}{C_{{0}}}^{3} ) {{\it r_0}}^{10} \\ &&
-4096\,{\alpha}^{6
}{{\it \ell}}^{2}{p}^{2}\sqrt {C_{{0}}} ( -35976\,\alpha\,{{\it 
\ell}}^{6}{p}^{6}C_{{0}}+163840\,{\alpha}^{4}{p}^{8}{C_{{0}}}^{4}-515\,
{{\it \ell}}^{8}{p}^{6} \\ &&
-3186688\,{p}^{4}{{\it \ell}}^{2}{\alpha}^{3}{C_{
{0}}}^{3}-123392\,{\alpha}^{3}{{\it \ell}}^{2}{p}^{8}{C_{{0}}}^{3}-
401248\,{p}^{4}{{\it \ell}}^{6}\alpha\,C_{{0}}+2848\,{\alpha}^{2}{{\it 
\ell}}^{4}{p}^{8}{C_{{0}}}^{2} \\ &&
-389120\,{p}^{6}{{\it \ell}}^{2}{\alpha}^{
3}{C_{{0}}}^{3}+1894912\,{p}^{4}{{\it \ell}}^{4}{\alpha}^{2}{C_{{0}}}^{
2}+24576\,{p}^{2}{{\it \ell}}^{4}{\alpha}^{2}{C_{{0}}}^{2}+30780\,{{
\it \ell}}^{8}{p}^{4} \\ &&
+1004800\,{\alpha}^{2}{{\it \ell}}^{4}{p}^{6}{C_{{0
}}}^{2}+442368\,{{\it \ell}}^{6}\alpha\,C_{{0}}+48640\,{{\it \ell}}^{8}+
851968\,{p}^{6}{\alpha}^{4}{C_{{0}}}^{4}-10656\,{p}^{2}{{\it \ell}}^{8}
 \\ &&
 -1641984\,{p}^{2}{{\it \ell}}^{6}\alpha\,C_{{0}} ) {{\it r_0}}^{9}
-16384\,{\alpha}^{6}{{\it \ell}}^{3}p ( -320\,{p}^{10}{{\it \ell}}^
{2}{\alpha}^{3}{C_{{0}}}^{3}+4096\,{p}^{10}{\alpha}^{4}{C_{{0}}}^{4}- \\ &&
1766400\,{p}^{6}{{\it \ell}}^{2}{\alpha}^{3}{C_{{0}}}^{3}-290816\,{p}^{
2}{{\it \ell}}^{6}\alpha\,C_{{0}}+2435072\,{p}^{4}{{\it \ell}}^{4}{
\alpha}^{2}{C_{{0}}}^{2}-2048\,{{\it \ell}}^{8} \\ &&
-61792\,\alpha\,{{\it 
\ell}}^{6}{p}^{6}C_{{0}}+706\,{p}^{8}{{\it \ell}}^{6}\alpha\,C_{{0}}+
90592\,{\alpha}^{2}{{\it \ell}}^{4}{p}^{6}{C_{{0}}}^{2}-417792\,{p}^{2}
{{\it \ell}}^{4}{\alpha}^{2}{C_{{0}}}^{2} \\ &&
+19832\,{\alpha}^{2}{{\it \ell}
}^{4}{p}^{8}{C_{{0}}}^{2}-1152\,{p}^{2}{{\it \ell}}^{8}+7080\,{{\it \ell
}}^{8}{p}^{4}-281\,{{\it \ell}}^{8}{p}^{6}+146016\,{p}^{4}{{\it \ell}}^{
6}\alpha\,C_{{0}} \\ &&
+245760\,{\alpha}^{4}{p}^{8}{C_{{0}}}^{4}-245504\,{
\alpha}^{3}{{\it \ell}}^{2}{p}^{8}{C_{{0}}}^{3}-49152\,{p}^{4}{{\it \ell
}}^{2}{\alpha}^{3}{C_{{0}}}^{3}+565248\,{p}^{6}{\alpha}^{4}{C_{{0}}}^{
4} ) {{\it r_0}}^{8} \\ &&
-32768\,{\alpha}^{7}{{\it \ell}}^{4}{p}^{2}
\sqrt {C_{{0}}} ( 32768\,{p}^{4}{\alpha}^{3}{C_{{0}}}^{3}+393216
\,{p}^{2}{{\it \ell}}^{2}{C_{{0}}}^{2}{\alpha}^{2}-988\,\alpha\,{{\it 
\ell}}^{4}{p}^{8}C_{{0}} \\ &&
-5248\,{\alpha}^{2}{{\it \ell}}^{2}{p}^{8}{C_{{0
}}}^{2}+84352\,{{\it \ell}}^{6}{p}^{2}+30720\,{\alpha}^{3}{p}^{8}{C_{{0
}}}^{3}+112224\,\alpha\,{{\it \ell}}^{4}{p}^{6}C_{{0}}-60284\,{{\it \ell
}}^{6}{p}^{4} \\ &&
+223232\,{\alpha}^{2}{{\it \ell}}^{2}{p}^{6}{C_{{0}}}^{2}+
1543\,{p}^{6}{{\it \ell}}^{6}-3501056\,{\alpha}^{2}{{\it \ell}}^{2}{p}^{
4}{C_{{0}}}^{2}-672768\,\alpha\,{{\it \ell}}^{4}{p}^{4}C_{{0}} \\ &&
+1208320
\,{p}^{2}\alpha\,{{\it \ell}}^{4}C_{{0}}+1044480\,{\alpha}^{3}{p}^{6}{C
_{{0}}}^{3}+7168\,{{\it \ell}}^{6} ) {{\it r_0}}^{7} \\ &&
+16384\,{
\alpha}^{7}{{\it \ell}}^{5}{p}^{3} ( 371968\,{\alpha}^{2}{{\it \ell
}}^{2}{p}^{6}{C_{{0}}}^{2}+10352\,\alpha\,{{\it \ell}}^{4}{p}^{6}C_{{0}
}-31744\,{{\it \ell}}^{6}+8568832\,{p}^{2}{{\it \ell}}^{2}{C_{{0}}}^{2}{
\alpha}^{2} \\ &&
+866304\,{\alpha}^{3}{p}^{6}{C_{{0}}}^{3}-2304\,{\alpha}^{2
}{{\it \ell}}^{2}{p}^{8}{C_{{0}}}^{2}-4709376\,{\alpha}^{2}{{\it \ell}}^
{2}{p}^{4}{C_{{0}}}^{2}-712864\,\alpha\,{{\it \ell}}^{4}{p}^{4}C_{{0}}+
 \\ &&
 2083072\,{p}^{2}\alpha\,{{\it \ell}}^{4}C_{{0}}+589824\,{p}^{2}{\alpha}
^{3}{C_{{0}}}^{3}+66112\,{{\it \ell}}^{6}{p}^{2}-7757824\,{p}^{4}{
\alpha}^{3}{C_{{0}}}^{3}-2716\,{{\it \ell}}^{6}{p}^{4} \\ &&
-294912\,{{\it 
\ell}}^{4}C_{{0}}\alpha+99\,{p}^{6}{{\it \ell}}^{6}+11264\,{\alpha}^{3}{
p}^{8}{C_{{0}}}^{3} ) {{\it r_0}}^{6} \\ &&
-65536\,{\alpha}^{8}{{\it 
\ell}}^{6}{p}^{4}\sqrt {C_{{0}}} ( -3864\,{p}^{4}{{\it \ell}}^{4}+
59904\,{C_{{0}}}^{2}{p}^{6}{\alpha}^{2}-1431552\,{p}^{4}{C_{{0}}}^{2}{
\alpha}^{2}+203680\,{{\it \ell}}^{4}{p}^{2} \\ &&
+2169344\,{{\it \ell}}^{2}{p}
^{2}C_{{0}}\alpha-237056\,{{\it \ell}}^{4}+2760704\,{p}^{2}{C_{{0}}}^{2
}{\alpha}^{2}+2432\,C_{{0}}{p}^{6}{{\it \ell}}^{2}\alpha+243\,{p}^{6}{{
\it \ell}}^{4} \\ &&
-441088\,\alpha\,{{\it \ell}}^{2}{p}^{4}C_{{0}}-630784\,{{
\it \ell}}^{2}C_{{0}}\alpha ) {{\it r_0}}^{5} \\ &&
-131072\,{\alpha}^{8}
{{\it \ell}}^{7}{p}^{3} ( 384\,{p}^{4}{{\it \ell}}^{4}-1471488\,{p}
^{4}{C_{{0}}}^{2}{\alpha}^{2}+622592\,{p}^{2}{C_{{0}}}^{2}{\alpha}^{2}
+315\,{p}^{6}{{\it \ell}}^{4} \\ &&
-180\,{p}^{8}{{\it \ell}}^{2}\alpha\,C_{{0}
}-378240\,\alpha\,{{\it \ell}}^{2}{p}^{4}C_{{0}}+40192\,{{\it \ell}}^{4}
{p}^{2}+1600\,C_{{0}}{p}^{6}{{\it \ell}}^{2}\alpha+199424\,{C_{{0}}}^{2
}{p}^{6}{\alpha}^{2} \\ &&
-20480\,{{\it \ell}}^{4}+757760\,{{\it \ell}}^{2}{p}
^{2}C_{{0}}\alpha ) {{\it r_0}}^{4}-1048576\,{\alpha}^{9}{{\it 
\ell}}^{8}{p}^{4}\sqrt {C_{{0}}} ( 31232\,{{\it \ell}}^{2}-824\,{{
\it \ell}}^{2}{p}^{4} \\ &&
-159744\,{p}^{2}C_{{0}}\alpha+544\,\alpha\,{p}^{6}
C_{{0}}-171\,{p}^{6}{{\it \ell}}^{2}+54912\,\alpha\,{p}^{4}C_{{0}}-
34464\,{{\it \ell}}^{2}{p}^{2} ) {{\it r_0}}^{3} \\ &&
-4194304\,{\alpha}
^{9}{{\it \ell}}^{9}{p}^{5} ( -63\,{{\it \ell}}^{2}{p}^{4}-452\,{{
\it \ell}}^{2}{p}^{2}-3072\,{{\it \ell}}^{2}+18\,\alpha\,{p}^{6}C_{{0}}+
704\,\alpha\,{p}^{4}C_{{0}} \\ &&
+13600\,{p}^{2}C_{{0}}\alpha-14080\,\alpha
\,C_{{0}} ) {{\it r_0}}^{2}-33554432\,{\alpha}^{10}{{\it \ell}}^{
10}{p}^{6}\sqrt {C_{{0}}} \left( 720+148\,{p}^{2}+9\,{p}^{4} \right) {
\it r_0} \\ &&
-2684354560\,{\alpha}^{10}{{\it \ell}}^{11}{p}^{7}-301989888\,{
\alpha}^{10}{{\it \ell}}^{11}{p}^{9} \\
\\
C^{\text{D}}_2 &=&
3\,{{\it r_0}}^{5}\sqrt {C_{{0}}} \left( {{\it r_0}}^{2}+16\,\alpha
 \right) ^{2} \left( 4\,{\it r_0}\,\alpha\,p\sqrt {C_{{0}}}+8\,\alpha\,
{\it \ell}-{\it \ell}\,{{\it r_0}}^{2} \right) ^{2}{p}^{2} ( 64\,{
\alpha}^{2}{\it \ell}\,p\sqrt {C_{{0}}}{\it r_0} \\ &&
+16\,{p}^{2}C_{{0}}{
\alpha}^{2}{{\it r_0}}^{2}-16\,{\alpha}^{2}{{\it \ell}}^{2}{p}^{2}-16\,{
{\it r_0}}^{2}{{\it \ell}}^{2}\alpha-8\,{\it \ell}\,{{\it r_0}}^{3}\alpha
\,p\sqrt {C_{{0}}}+{{\it r_0}}^{4}{{\it \ell}}^{2} ) ^{3}
\end{eqnarray*}
}

\bibliographystyle{unsrt}
\bibliography{EH5D_EGB_references}

\end{document}